\documentclass[sigconf, authorversion, noacm]{acmart}
\renewcommand\footnotetextcopyrightpermission[1]{}

\AtBeginDocument{%
  }

\setcopyright{acmlicensed}
\copyrightyear{2026}
\acmYear{2026}
\acmDOI{XXXXXXX.XXXXXXX}
\acmConference[Conference acronym 'XX]{Make sure to enter the correct
  conference title from your rights confirmation email}{June 03--05,
  2018}{Woodstock, NY}
\acmISBN{978-1-4503-XXXX-X/2018/06}

\usepackage{microtype}
\usepackage{graphicx}
\usepackage{subfigure} 
\usepackage{booktabs} 
\usepackage{tabularx}
\usepackage{enumitem}
\usepackage{algorithm}
\usepackage{algorithmic}
\usepackage{amsmath}
\usepackage{mathtools}
\usepackage{amsthm}
\usepackage[capitalize,noabbrev]{cleveref}
\usepackage[textsize=tiny]{todonotes}
\usepackage{xcolor}
\usepackage{xspace}
\usepackage{multirow}
\usepackage{soul}

\usepackage[most]{tcolorbox}
\definecolor{promptbgcolor}{RGB}{245,245,245} 
\definecolor{promptframecolor}{RGB}{180,180,180} 

\newtcolorbox{promptbox}[1][]{
  enhanced,
  boxrule=0.5pt,
  colback=promptbgcolor,
  colframe=promptframecolor,
  fonttitle=\bfseries,
  coltitle=black,
  attach boxed title to top left={yshift=-2mm, xshift=2mm},
  boxed title style={
    boxrule=0pt,
    colframe=white,
    colback=white,
  },
  title={#1},
  arc=2mm, 
  breakable, 
}

\newcommand{\ourmethod}{AgenticTagger\xspace}
\newcommand{\ourmethodauto}{AgenticTagger\xspace}

\begin{document}
\title{\ourmethodauto: Generating Structured Item Representation for Recommendation with LLM Agents}

\author{Zhouhang Xie}
\authornote{Equal contribution.}
\authornote{Work done as a Student Researcher at Google.}
\affiliation{
  \institution{UC San Diego}
  \city{La Jolla}
  \state{CA}
  \country{USA}}
\email{zhx022@ucsd.edu}

\author{Bo Peng}
\authornotemark[1]
\affiliation{
  \institution{Google}
  \city{Mountain View}
  \state{CA}
  \country{USA}}
\email{bopen@google.com}

\author{Zhankui He}
\authornotemark[1]
\affiliation{
  \institution{Google DeepMind}
  \city{Mountain View}
  \state{CA}
  \country{USA}}
\email{zhankui@google.com}

\author{Ziqi Chen}
\authornotemark[1]
\affiliation{
  \institution{Google}
  \city{Mountain View}
  \state{CA}
  \country{USA}}
\email{zqchen@google.com}

\author{Alice Han}
\affiliation{
  \institution{Google}
  \city{Mountain View}
  \state{CA}
  \country{USA}}
\email{superalice@google.com}

\author{Isabella Ye}
\affiliation{
  \institution{Google}
  \city{Mountain View}
  \state{CA}
  \country{USA}}
\email{isabellasye@google.com}

\author{Benjamin Coleman}
\affiliation{
  \institution{Google DeepMind}
  \city{Mountain View}
  \state{CA}
  \country{USA}}
\email{colemanben@google.com}

\author{Noveen Sachdeva}
\affiliation{
  \institution{Google DeepMind}
  \city{Mountain View}
  \state{CA}
  \country{USA}}
\email{noveen@google.com}

\author{Fernando Pereira}
\affiliation{
  \institution{Google DeepMind}
  \city{Mountain View}
  \state{CA}
  \country{USA}}
\email{pereira@google.com}

\author{Julian McAuley}
\affiliation{
  \institution{UC San Diego}
  \city{La Jolla}
  \state{CA}
  \country{USA}}
\email{jmcauley@ucsd.edu}

\author{Wang-Cheng Kang}
\affiliation{
  \institution{Google DeepMind}
  \city{Mountain View}
  \state{CA}
  \country{USA}}
\email{wckang@google.com}

\author{Derek Zhiyuan Cheng}
\affiliation{
  \institution{Google DeepMind}
  \city{Mountain View}
  \state{CA}
  \country{USA}}
\email{zcheng@google.com}

\author{Beidou Wang}
\affiliation{
  \institution{Google}
  \city{Mountain View}
  \state{CA}
  \country{USA}}
\email{beidou@google.com}

\author{Randolph Brown}
\affiliation{
  \institution{Google}
  \city{Mountain View}
  \state{CA}
  \country{USA}}
\email{rgb@google.com}

\renewcommand{\shortauthors}{Xie et al.}

\begin{abstract}
High-quality representations are a core requirement for effective recommendation.
In this work, we study the problem of LLM-based descriptor generation, i.e., keyphrase-like natural language item representation generation frameworks with minimal constraints on downstream applications.
We propose \ourmethodauto, a framework that queries LLMs for representing items with sequences of text descriptors.
However, open-ended generation provides little control over the generation space,  leading to high cardinality, low-performance descriptors that render downstream modeling challenging. 
To this end, \ourmethodauto features two core stages: (1) a vocabulary-building stage in which a set of hierarchical, low-cardinality, and high-quality descriptors is identified, and (2) a vocabulary-assignment stage in which LLMs assign in-vocabulary descriptors to items. 
To effectively and efficiently ground vocabulary in the item corpus of interest, we design a multi-agent reflection mechanism in which an architect LLM iteratively refines the vocabulary guided by parallelized feedback from annotator LLMs that validate the vocabulary against item data.
Experiments on public and private data show \ourmethodauto brings consistent improvements across diverse recommendation scenarios, including generative and term-based retrieval, ranking, and controllability-oriented, critique-based recommendation.
\end{abstract}






\maketitle

\begin{figure}[t]
    \centering
    \includegraphics[width=0.44\textwidth]{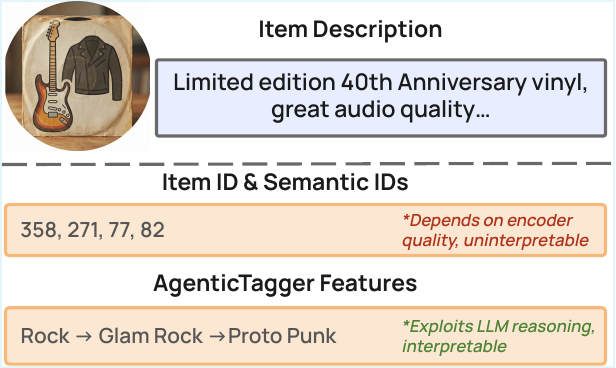}
    \caption{Illustration of the proposed method. Items are represented using an \textit{ordered sequence} of LLM-generated descriptors that compress task-relevant information.}
    \label{fig:motivation}
    \vspace{-1em}
\end{figure}

\section{Introduction}
\label{introduction}

High-quality item representations are the foundation of effective recommendation systems~\cite{Steffen2009BPR, kang2021learning, rajput2023recommender, Anima2024Better, hou2023vqrec, coleman2023unified, Zhang2024NoteLLM, Zhang2025NoteLLM2}. 
These representations are commonly derived from collaborative filtering~\cite{Steffen2009BPR, Yehuda2009MF, Santosh2013FISM, he2017neuralcollaborativefiltering, Kang2018SelfAttentiveSR} and/or content signals~\cite{sharma2015featurebased, mooney2000content, rajput2023recommender, Anima2024Better, hou2023vqrec, hou2025actionpiece, hou2025rpg, wang2024colla} via representation learning.
Recently, there have been numerous attempts to enhance item representations with LLMs, typically by generating auxiliary unstructured item descriptions~\cite{valizadeh2025languagemodelssemanticaugmenters, Ren2024EnhancingSeq, shi2025mattersllmbasedfeatureextractor, yada2024newsrecommendationcategorydescription, wang2024llmasda} or prompting LLMs to infer user interests from interacted items and directly make recommendations~\cite{liu2025inferencecomputationscalingfeature, liu2025improvingLLMPoweredRec, jalan2024llmbrec, jiang-etal-2025-reclm, wang2024llmDA}.
However, these approaches either require modeling high-cardinality unstructured natural language descriptions or deploying specialized recommendation frameworks, whereas large-scale recommender systems frequently benefit from low-cardinality, broadly applicable item representations such as hashed categorical features~\cite{kang2021learning, coleman2023unified, Anima2024Better, he2025plumadaptingpretrainedlanguage, hou2023vqrec, hou2025actionpiece, hou2025rpg}.
Thus, despite recent advances in both LLM-enhanced and structured item representations, it is unclear how to effectively leverage LLMs' strong semantic understanding ability to enhance structured item representations.

\begin{figure*}[h]
    \centering
    \includegraphics[width=0.90\textwidth]{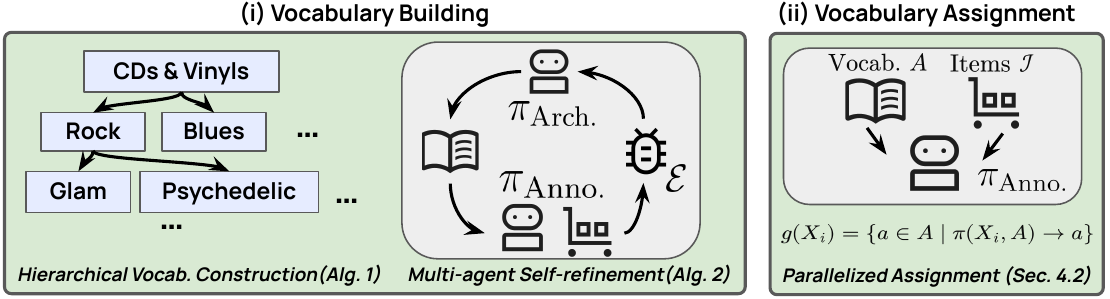}
    \caption{\textbf{The \ourmethod Framework.} \ourmethodauto exploits LLMs' content-understanding ability to automatically construct an interpretable, hierarchical descriptor vocabulary from data. Then, relevant descriptors in the vocabulary are assigned to corresponding items by querying LLMs in parallel over the item corpus. These features could then power various downstream applications, such as generative retrieval, ranking, and critique-based recommendation.}
    \label{fig:llmtagger_main_fig}
\end{figure*}

To this end, we study item-side, LLM-based generation frameworks for discrete feature representations (i.e., tag-like natural language descriptors), capable of exploiting LLMs' reasoning ability to produce compact, structured, and fixed-slot nominal features with minimal constraints on downstream models, as shown in~\Cref{fig:motivation}.
We argue that this setting holds three complementary values compared to its alternatives, such as generating unstructured item descriptions with LLMs or prompting LLMs directly for recommendations.
First, discrete features such as hashed item IDs and text descriptors provide irreplaceable value in recommender systems, due to their compatibility with diverse architectures~\cite{Zhang2024NoteLLM, Zhang2025NoteLLM2, wang2017DCN, Wang2021DCNV2, Kabbur2013FISM, cheng2016widedeep, song2019AutoInt}, their effectiveness in improving recommendation accuracy~\cite{rajput2023recommender, hou2023vqrec, hou2025actionpiece, Anima2024Better}, and their scalability when deployed in keyword-based retrieval or advertisement targeting~\cite{christian2010selecting, google2026aboutkeywords}.
Second, item representation scales easily in the typical case where a recommender system serves more users than items, and the item corpus remains more stable than user interests~\cite{McAuley2015ImageBasedRec, Gao2022KuaiRec}. 
Finally, as we discuss in a later section (\Cref{subsec:interpretability_and_controllability}), text-based descriptor representations provide inherent interpretability and are well-suited for user-facing applications.

While existing efforts typically directly generate unstructured item descriptions from LLMs~\cite{valizadeh2025languagemodelssemanticaugmenters, Ren2024EnhancingSeq, shi2025mattersllmbasedfeatureextractor, yada2024newsrecommendationcategorydescription, wang2024llmasda}, free-form LLM generations offer no control over their exact word choices, and thus there is no way to enforce structural constraints.
For example, it is challenging to enforce mutual exclusivity between descriptions from free-form generation (\Cref{subsec:free_form_tag_as_semid}), making these descriptors unsuitable for our use case.
Further, a good set of text descriptors should ideally capture meaningful frequent patterns in the item corpus~\cite{xie-etal-2025-latent, rajput2023recommender, Anima2024Better}.
Yet, recommender system corpora generally exceed the context limit of LLMs~\cite{McAuley2015ImageBasedRec, harper2016movielens}, and thus it is challenging to incorporate the abstract constraint of generating descriptors that meaningfully highlight item similarities and differences w.r.t. other items, as this requires a global view of the entire item corpus.
Finally, as LLMs can pick any arbitrary surface form (e.g., "Blues music" and "Blues great hits") for the same underlying concept in free-form generation, such an approach often leads to vocabulary explosion, where most generated descriptors occur only once across the item corpus, as we discuss in~\Cref{subsec:vocab_ultilization}.
This renders many features unusable without further processing, compromising downstream recommendation performance (\Cref{subsec:free_form_tag_as_semid}).

To address these issues, we propose \ourmethod, a multi-LLM framework that queries LLMs for natural language descriptors (i.e., tag-like descriptors shown in~\Cref{fig:motivation}), where the descriptors come from an automatically learned low-cardinality, hierarchically structured vocabulary.
Specifically, \ourmethod features two key stages: (1) \textbf{vocabulary building} (\Cref{subsec:vocab_building}), where \ourmethodauto automatically learns a hierarchical coarse-to-fine descriptor taxonomy, drawing insights from LLMs for taxonomy mining~\cite{Wan2024TnTLLM, xie-etal-2025-latent, pham-etal-2024-topicgpt} and codebook learning for item representation in recommendation~\cite{hou2023vqrec, hou2025actionpiece, hou2025rpg, Anima2024Better, he2025plumadaptingpretrainedlanguage, rajput2023recommender}; and (2) \textbf{vocabulary assignment} (\Cref{subsec:vocab_assignment}), which prompts LLMs to assign relevant descriptors from the vocabulary to items, constraining their generation space.
By mining a high-quality, low-cardinality descriptor vocabulary and performing in-vocabulary generation, \ourmethod effectively constrains the LLMs' generation space while equipping LLMs with a global, hierarchical view of the item corpus, enforcing structural constraints on LLM-generated descriptors.

Under the general \ourmethodauto framework, a core design challenge is how to effectively ``propagate'' signals through LLM generations during the vocabulary-building stage to improve the quality of the vocabulary and better cover diverse items in the item corpus.
To this end, following the numerous recent successes in self-improving LLM agents through verbal feedback~\cite{shinn2023reflection, majumder2024clin, xie-etal-2024-shot-dialogue}, we adopt an LLM-based refinement loop, which refines the vocabulary by aiming for better coverage of the item corpus through LLM-generated failure reports on items that do not fit into the current vocabulary.
Unlike these prior works, \ourmethodauto uniquely adopts a multi-agent architect-annotator framework, where a stronger (yet slower) architect LLM maintains the descriptor vocabulary, while numerous copies of the weaker (yet faster) annotator LLM scan through the item corpus in parallel, making \ourmethodauto scalable.

We conduct experiments on \ourmethod across public benchmarks and a private dataset, showing the broad applicability of \ourmethod via its improvements to diverse recommendation settings, including generative recommendation and ranking.
We also show that these natural language descriptors natively bring interpretability and controllability to recommender systems, such as by making generative recommender models compatible with critiques by controlling their decoding process via user feedback.
Our contributions are as follows:
\begin{itemize}[nosep, leftmargin=*]
    \item To the best of our knowledge, we are the first to apply agentic frameworks to feature mining to benefit recommender systems, opening up a new direction for future work.
    \item We identify a core challenge in LLM-based structured item representation generation, namely, the lack of vocabulary control, and propose a multi-agent, two-stage descriptor generation framework to address this issue.
    \item We show the value of LLM-agent-produced, structured natural language item representations in recommender systems, for their performance, interpretability, and flexibility.
\end{itemize}

\section{Related Work}

\noindent\textbf{LLM-generated Features in RecSys.} Prior work in introducing LLM-produced features in recommender systems broadly falls into two streams: LLM-as-encoders~\cite{Zhang2024NoteLLM, Zhang2025NoteLLM2} and LLM-as-text-generators~\cite{Ren2024EnhancingSeq, yada2024newsrecommendationcategorydescription, valizadeh2025languagemodelssemanticaugmenters, liu2024Once, liu2025inferencecomputationscalingfeature}, and this work aims at enhancing the latter. 
In contrast to the popular goal of generating user profiles that improve an LLM itself at the recommendation task~\cite{liu2025inferencecomputationscalingfeature, liu2025improvingLLMPoweredRec}, we consider specifically the case where the recommendation model is a general feature-aware ML model, which aligns well with most existing deployed recommendation models.
Compared to the most similar works that encode LLM-produced descriptions~\cite{Ren2024EnhancingSeq, yada2024newsrecommendationcategorydescription, valizadeh2025languagemodelssemanticaugmenters}, we are the first to study fixed-slot, nominal feature generation, and the first to demonstrate the value of LLM agents in feature generation for RecSys.\\

\noindent\textbf{Discrete Representation in RecSys.} 
Our work is also similar to a series of efforts in building better discrete features for RecSys, which typically involve discretizing item representations with product quantization and its variants~\cite{Jegou2011ProductQ, Ge2014OptimizedPQ, hou2023vqrec, hou2025rpg}, RQ-VAE~\cite{van2017neuraldiscrete, rajput2023recommender}, and K-means~\cite{ju2025grid, Zhang2024RecGPTGP}.
However, these methods do not leverage the reasoning ability of LLMs. 
As we will later show in the paper, discrete features produced from raw text by exploiting LLMs' content understanding ability outperform their counterparts, such as RQ-VAE-based semantic IDs.\\

\noindent\textbf{LLM Agents for RecSys.} 
There have been numerous recent efforts in leveraging LLM agents to improve RecSys. These frameworks generally focus on leveraging LLMs for item recommendation~\cite{shu2023rahrecsysassistanthumanhumancenteredrecommendation, Zhao2024LetMe, thakkar2024personalizedrecommendationsystemsusing, wang2024recmind, xia2026multiagentcollaborativefilteringorchestrating, huang2025recaiagent}, or enabling better interactivity by incorporating LLMs and LLM agents into the recommendation-serving stage~\cite{friedman2023leveraginglargelanguagemodels, He2023Large, zhang2024prospectpersonalizedrecommendationlarge}.
However, these frameworks typically require specialized pipelines, whereas \ourmethod focuses on feature generation, which potentially benefits diverse downstream recommendation models. 
Meanwhile, to the best of our knowledge, \ourmethodauto is the first framework to study agentic feature generation for recommender systems.

\section{Problem Statement}

Consider the standard recommendation setting with item features. 
Given a set of users $\mathcal{U}$ and items $\mathcal{I}$, the objective is to predict unobserved items of interest for a user $u \in \mathcal{U}$ based on their history of interactions $S_u \subset \mathcal{I}$. 
Each item $i \in \mathcal{I}$ is associated with raw textual attributes $X_i \in \mathcal{X}$, where $\mathcal{X}$ is the original feature space.
In the context of feature generation, our goal is to find an optimal one-to-many transformation function, $g: \mathcal{X} \to \mathcal{P}(A)$, that maps these original attributes to a new discrete feature space $A$. The optimal transformation, $g^*$, is one that maximizes the performance of a downstream recommender system, $f$:
$$
g^* = \arg\max_{g} \mathcal{M}(f_{\theta^*(g)})
$$
where $f_{\theta^*(g)}$ is the recommender model trained using the generated features, and $\mathcal{M}$ represents relevant performance metrics, e.g., offline metrics and key business metrics. 
Note that while \ourmethod does not directly optimize for end performance due to the challenge of propagating gradients through an API-supported LLM, the best vocabulary should ideally, in our problem formulation, maximize downstream task performance.
We provide an overview of the notations used in~\Cref{tab:notations}.

\section{\ourmethodauto}
\label{sec:our_method_advanced}

As shown in~\Cref{fig:llmtagger_main_fig}, the overall pipeline of \ourmethodauto involves two key stages: the vocabulary-building stage, where we produce a descriptor vocabulary by observing typical items in the dataset, and the vocabulary-assignment stage, where we assign the best set of descriptors to each item.

\subsection{Vocabulary Building}
\label{subsec:vocab_building}

\begin{algorithm}[tb]
   \caption{Vocabulary Building}
   \label{alg:hierarchical}
\begin{algorithmic}
   \STATE {\bfseries Input:} Item corpus $\mathcal{I}$, LLM Ensemble $\Pi$
   \STATE {\bfseries Hyperparameters:} Max hierarchy depth $D_{max}$, split threshold $\tau_{split}$
   \STATE {\bfseries Output:} Hierarchical descriptor set $A_{all}$
   \STATE
   \STATE $A_{all} \gets \emptyset$
   \STATE $p_0 \gets \text{new Descriptor}(\text{items} \gets \mathcal{I}, \text{parent} \gets \text{null})$
   \STATE $Q \gets \{ p_0 \}$
   \FOR{{\bfseries each} $d \in \{1, \dots, D_{max}\}$}
      \IF{$|Q| = 0$} \STATE \textbf{break} \ENDIF
      \STATE $Q_{next} \gets \emptyset$
      \FOR{{\bfseries each} parent feature $p \in Q$}
         \IF{$|p.\text{items}| \geq \tau_{split}$}
            \STATE $A_{children} \gets \text{Algorithm \ref{alg:refinement}}(p.\text{items}, \pi_{\text{Annotator}}, p)$
            \STATE $A_{all} \gets A_{all} \cup A_{children}$
            \STATE $Q_{next} \gets Q_{next} \cup A_{children}$
         \ENDIF
      \ENDFOR
      \STATE $Q \gets Q_{next}$
   \ENDFOR
   \STATE {\bfseries return} $A_{all}$
\end{algorithmic}
\end{algorithm}

During the vocabulary-building stage, \ourmethodauto automatically learns a vocabulary from the item corpus.
However, gradient-based vocabulary learning~\cite{rajput2023recommender, hou2023vqrec, hou2025rpg, Zhang2024RecGPTGP} is inapplicable to prompting-based LLM agents, and thus the common solution is prompting an LLM in mini-batches~\cite{Wan2024TnTLLM, pham-etal-2024-topicgpt} and refining an in-context vocabulary via LLM self-refinement~\cite{madaan2023selfrefine}.
Yet, such an iterative approach involves a non-parallelizable loop over the whole item corpus, as LLMs must sequentially refine their in-context vocabulary after seeing each batch of items, which scales poorly to large datasets~\cite{xie-etal-2025-latent}.

\subsubsection{Hierarchical Vocabulary Construction.} In light of the above issues, in~\ourmethodauto, we opt to scale existing LLM-based frameworks by making the vocabulary-building process parallelizable, where multiple LLMs are deployed concurrently to update a shared vocabulary that describes observed items.
An overview of this process is illustrated in~\Cref{alg:hierarchical} and~\Cref{alg:refinement}.
Specifically, the iterative feature generation follows a hierarchical-clustering-like pipeline (\Cref{alg:hierarchical}), where we generate an initial set of features that partitions the whole item corpus, followed by a top-down procedure to generate finer-grained features that further partition items associated with each feature into smaller subsets. 
For example, if the initial features include high-level category information such as ``CDs and Vinyl'', all relevant items will then form a subcluster, for generating finer-grained features such as ``Rock'' and ``Blues'', as shown in~\Cref{fig:llmtagger_main_fig}-(i).
By decomposing the task of generating an entire vocabulary into a top-down hierarchical procedure, \ourmethod effectively limits the number of descriptors an LLM needs to produce at a time, which scales better to a large corpus. Meanwhile, the resulting hierarchical vocabulary naturally organizes items in a coarse-to-fine taxonomy, similar to hierarchical semantic IDs~\cite{rajput2023recommender, he2025plumadaptingpretrainedlanguage}, which are known to act as effective item representations.

\subsubsection{Multi-agent Self-refinement.} At the core of this top-down generation loop is a multi-agent iterative refinement (\Cref{fig:llmtagger_main_fig}-(ii) and~\Cref{alg:refinement}) subroutine reused for each feature generation job $Q_{i}$, which aims to produce a set of features that describe common items in the corresponding subset.
Our core insight is leveraging a stronger (though potentially slower) LLM, $\pi_{Architect}$, to generate and refine the vocabulary subset relevant to each $Q_i$, and copies of a faster (and potentially weaker) LLM, $\pi_{Annotator}$, which attempt to annotate the current item corpus using the available vocabulary and provide feedback to $\pi_{Architect}$ to improve the vocabulary.
Unlike in single-LLM self-refinement~\cite{Wan2024TnTLLM, pham-etal-2024-topicgpt, madaan2023selfrefine}, the most costly $\pi_{Annotator}$ operations are parallelized in \ourmethod, while we can still use their feedback to prompt $\pi_{Architect}$ to refine the vocabulary.

As shown in~\Cref{alg:refinement}, during the subroutine execution, we first prompt the $\pi_{\text{Architect}}$ LLM to generate the initial set of descriptors by observing a small subset of items, $\text{Distill}(I_{sub})$, in its context, and then deploy copies of (or batched queries to) the annotator LLMs $\pi_{\text{Annotator}}$ to determine if any of the currently known features apply to each item in $I_{sub}$, generating a set of error suggestions $\mathcal{E}$. 
The specific prompts are as shown in~\Cref{subsec:sample_prompts_sports}.
The error reports $\mathcal{E}$ then act as inductive biases for prompting the vocabulary maintainer LLM $\pi_{\text{Architect}}$ to refine the current set of features.
To avoid overcrowding $\pi_{\text{Architect}}$'s context, $\text{Distill}(\cdot)$ refers to a general operation that selects diverse representative samples from a larger set of item descriptions, such as the centroids from K-Medoids clustering~\cite{Kaufman1987ClusteringBM}, as we use in our experiments. 
We repeat this iterative refinement loop until a fixed number of optimization steps has been reached, or a desired number of items in $I_{sub}$ can be described by the generated features $A$.
The union of features generated for each subset of the item corpus then forms the final feature space $A_{all}$, providing a coarse-to-fine description for the whole item corpus.

\begin{algorithm}[tb]
   \caption{Multi-agent Self-refinement}
   \label{alg:refinement}
\begin{algorithmic}
   \STATE {\bfseries Input:} Item corpus subset $I_{sub} \subseteq \mathcal{I}$; \\\hspace{2.7em} LLMs $\Pi = \{\pi_{\text{Architect}}, \pi_{\text{Annotator}}\}$
   \STATE {\bfseries Hyperparameters:} Max refinement cycles $C_{max}$, anomaly threshold $\tau_{anom}$
   \STATE {\bfseries Output:} Descriptor vocabulary $A$
   \STATE
   \STATE $A \gets \pi_{\text{Architect}}(\text{Distill}(I_{sub}))$
   \FOR{$c=1$ {\bfseries to} $C_{max}$}
      \STATE $(I_{assigned}, I_{unassigned}, \mathcal{E}) \gets \text{ParallelAssign}(I_{sub}, A, \pi_{\text{Annotator}})$
      \IF{$|I_{unassigned}| < \tau_{anom}$}
         \STATE \textbf{break}
      \ENDIF
      \STATE $A \gets \pi_{\text{Architect}}(A, \text{Distill}(\mathcal{E}))$
   \ENDFOR
   \STATE {\bfseries return} $A$
\end{algorithmic}
\end{algorithm}

\subsection{Vocabulary Assignment}
\label{subsec:vocab_assignment}

Once the feature vocabulary $A_{all}$ is built, the vocabulary assignment stage for \ourmethodauto is straightforward, where we simply constrain the descriptors output by the LLMs to fall into the vocabulary by prompting, i.e., $g(X_i) = \{a \in A \mid \pi(X_i, A) \to a\}$ for any raw item attribute $X_i$.
For example, in \ourmethodauto, for the sake of implementation simplicity, we opt to reuse the same operation from building the vocabulary, $\text{ParallelAssign}(I, A_{all}, \pi_{\text{Annotator}})$, where an LLM is prompted to assign features to an item by observing all features in its context, exploiting modern LLMs' long-context and reasoning abilities~\cite{zhao2025surveylargelanguagemodels}. 
As we later show in~\Cref{sec:experiments}, the generated descriptors are indeed of sufficient quality to support good performance in downstream recommendation tasks.
We provide further analysis and qualitative examples in~\Cref{subsec:generated_tags}.

We note that in principle, there might be multiple descriptors of fit at each depth of the vocabulary hierarchy (e.g., a CD that is both ``Rock'' and ``Blues''). 
To this end, we simply instruct the LLM to select a single best descriptor per level, enforcing structural constraints while preserving item information.
As we later showcase in our experiments (\Cref{sec:experiments}), this is sufficient for good downstream recommendation performance, and we leave better collision-resolving methods to future work.

\subsection{Complexity Analysis}

\subsubsection{Vocabulary Generation.} In \ourmethodauto's automated vocabulary-building algorithm, the parallel calls to the annotator LLMs during the vocabulary creation phase dominate. 
In its most basic version, the algorithm will incur $O(C_{max} \cdot N \cdot b^{D_{max}})$ LLM calls, assuming an item will be assigned to an average of $b$ features at each depth level $d \in \{1, \dots, D_{max}\}$, which grows quickly as the depth of the feature hierarchy grows. 
However, the expressiveness of the feature hierarchy also grows exponentially w.r.t. the depth of the hierarchy, similar to other hierarchical discrete item representation methods~\cite{rajput2023recommender, Zhang2024RecGPTGP}, and thus a few layers are often sufficient; as we later show in our experiments, 3-6 layers of features are sufficient for good performance across public benchmarks (\Cref{tab:performance_comparison_public}).

Further, since the goal of vocabulary building is solely to generate a set of feature descriptors, we can also heuristically constrain the branching factor by limiting how many features an item is assigned to at each layer, such as randomly assigning an item to one of its associated features during the branching operation. 
In practice, we observe that a branching factor of 1 is also sufficient for good performance (\Cref{subsubsec:scaling_to_larger_dataset}), in which case the complexity w.r.t. LLM calls becomes $O(C_{max} \cdot N \cdot D_{max})$.
Finally, compared to prior LLM-based document categorization methods~\cite{Wan2024TnTLLM, pham-etal-2024-topicgpt}, since \ourmethodauto does not have sequential dependency between most of its dominant LLM operations (specifically, the $\pi_{\text{Annotator}}$ annotations), the overall execution time can be further reduced approximately linearly w.r.t. the number of concurrent threads in the $\text{ParallelAssign}$ operation.

\begin{table*}[t]
\centering
\caption{\ourmethodauto for Generative Recommendation on academic benchmarks. For each metric, the best result is in \textbf{bold} and the second-best is \underline{underlined}. Our model's performance is reported across 3 runs.
We provide additional comparison against crossing-based IDs and free-form generations in \Cref{fig:apc_spm_comparison} and \Cref{subsec:free_form_tag_as_semid}.}
\label{tab:performance_comparison_public}
\resizebox{\textwidth}{!}{%
\begin{tabular}{@{}ll cc ccc cccccc@{}}
\toprule
\multirow{2}{*}{Datasets} & \multirow{2}{*}{Metric} & \multicolumn{2}{c}{ID-based} & \multicolumn{3}{c}{Feature + ID} & \multicolumn{5}{c}{Generative} \\
\cmidrule(lr){3-4} \cmidrule(lr){5-7} \cmidrule(lr){8-12}
& & BERT4Rec & SASRec & FDSA & S³-Rec & VQ-Rec & LMIndexer & HSTU & P5-CID & TIGER & \ourmethodauto \\
\midrule
\multirow{4}{*}{Sports} 
& R@5  & 0.0115 & 0.0233 & 0.0182 & 0.0251 & 0.0181 & 0.0222 & 0.0258 & \underline{0.0287} & 0.0264 & \textbf{0.0299} \tiny{$\pm$0.0010} \\
& N@5  & 0.0075 & 0.0154 & 0.0122 & 0.0161 & 0.0132 & 0.0142 & 0.0165 & 0.0179 & \underline{0.0181} & \textbf{0.0194} \tiny{$\pm$0.0007} \\
& R@10 & 0.0191 & 0.0350 & 0.0288 & 0.0385 & 0.0251 & —      & 0.0414 & \underline{0.0426} & 0.0400 & \textbf{0.0474} \tiny{$\pm$0.0002} \\
& N@10 & 0.0099 & 0.0192 & 0.0156 & 0.0204 & 0.0154 & —      & 0.0215 & 0.0224 & \underline{0.0225} & \textbf{0.0251} \tiny{$\pm$0.0004} \\
\midrule
\multirow{4}{*}{Beauty} 
& R@5  & 0.0203 & 0.0387 & 0.0267 & 0.0387 & 0.0434 & 0.0415 & \underline{0.0469} & 0.0468 & 0.0454 & \textbf{0.0492} \tiny{$\pm$0.0007} \\
& N@5  & 0.0124 & 0.0249 & 0.0163 & 0.0244 & 0.0311 & 0.0262 & 0.0314 & 0.0315 & \underline{0.0321} & \textbf{0.0332} \tiny{$\pm$0.0008} \\
& R@10 & 0.0347 & 0.0605 & 0.0407 & 0.0647 & \textbf{0.0741} & —      & 0.0704 & 0.0701 & 0.0648 & \underline{0.0728} \tiny{$\pm$0.0010} \\
& N@10 & 0.0170 & 0.0318 & 0.0208 & 0.0327 & 0.0372 & —      & 0.0389 & \underline{0.0400} & 0.0384 & \textbf{0.0408} \tiny{$\pm$0.0008} \\
\midrule
\multirow{4}{*}{CDs}    
& R@5  & 0.0326 & 0.0351 & 0.0226 & 0.0213 & 0.0314 & —      & 0.0417 & \underline{0.0505} & 0.0492 & \textbf{0.0544} \tiny{$\pm$0.0005} \\
& N@5  & 0.0201 & 0.0177 & 0.0137 & 0.0130 & 0.0209 & —      & 0.0275 & 0.0326 & \underline{0.0329} & \textbf{0.0358} \tiny{$\pm$0.0002} \\
& R@10 & 0.0547 & 0.0619 & 0.0378 & 0.0375 & 0.0485 & —      & 0.0638 & \underline{0.0785} & 0.0748 & \textbf{0.0836} \tiny{$\pm$0.0003} \\
& N@10 & 0.0271 & 0.0263 & 0.0186 & 0.0182 & 0.0264 & —      & 0.0346 & \underline{0.0416} & 0.0411 & \textbf{0.0453} \tiny{$\pm$0.0002} \\
\bottomrule
\end{tabular}%
}
\end{table*}

\subsubsection{Vocabulary Assignment.} Once the vocabulary is built, the assignment stage simply involves looping through all items of interest, and the assignment loop induces $O(|\mathcal{I}|)$ LLM calls, i.e., the cost of querying $\pi_{Annotator}$ in a single-pass manner over the item corpus.
To this end, after the vocabulary is built, \ourmethodauto can easily scale to large item corpora with parallelized calls to LLMs.
Further, since the only difference across LLM calls during the assignment stage is the item description, the inference can, in principle, be further accelerated by infrastructure optimizations such as KV-Cache~\cite{pope2023efficient}, which we leave for future work.

\subsection{Modeling \ourmethod Features}

Once the process is complete,  \ourmethod assigns a fixed maximum number of descriptors to an item, where each ``slot'' represents a categorical feature. 
In this work, we opt to assume the \textit{best} potential value for each slot is nominal, i.e., mutually exclusive, although there could be more than one suitable value at the same slot for each item (e.g., a song that is both ``Rock'' and ``Blues''). 
These assigned descriptors also have an inherent order, where a preceding descriptor represents the broader, parent category of its child descriptors.
For example, for an item to fit the ``Glam Rock'' descriptor, its preceding, broader feature would be ``Rock''.
As we later show, this formulation allows these features to be natively used as item IDs, such as semantic IDs and ID-like ranking features (\Cref{sec:experiments}).

\section{Main Experiments}
\label{sec:experiments}
\label{subsec:agentic_for_genrec}

In this section, we showcase the performance of the~\ourmethod-produced descriptor features by deploying them as discrete ID features in ML models, in particular, generative recommendation (\Cref{subsec:generative_recommendation}) and ranking (\Cref{subsec:ranking}). We then discuss an interpretability-oriented application, namely critique-based generative recommendation (\Cref{subsec:interpretability_and_controllability}), before other ablation studies and analyses.

\subsection{Generative Recommendation}
\label{subsec:generative_recommendation}

\subsubsection{Datasets and Evaluation.} 

We conduct our experiments across three domains from the Amazon Reviews dataset~\cite{McAuley2015ImageBasedRec}: ``Sports and Outdoor'' (\textbf{Sports}), ``Beauty" (\textbf{Beauty}), and ``CDs and Vinyl" (\textbf{CDs}); the dataset statistics are shown in~\Cref{tab:dataset_stats}.
For the larger CDs domain only, we constrain \ourmethodauto's branching factor to 1 for scalability and apply the vanilla \ourmethodauto algorithm to the other two domains (see \Cref{subsubsec:scaling_to_larger_dataset} for discussions).
We adopt the commonly used ``last-out'' splitting~\cite{Kang2018SelfAttentiveSR, rajput2023recommender, hou2025rpg}, where we hold out the last chronologically interacted item for each user as the test set, and the second-last item as the validation set.
Following recent work~\cite{rajput2023recommender, hou2023vqrec, wang2024learnableitemtokenizationgenerative}, we evaluate the performance of models with Recall@$K$ and NDCG@$K$, where $K \in \{5, 10\}$.
To eliminate randomness in model optimization, we repeat our experiments with 3 random seeds and report the average metrics.
We use \texttt{Gemini-2.5-flash} and \texttt{Gemini-2.5-flash-lite} for the architect and annotator LLMs, respectively, in our main experiment.
We discuss results from \texttt{Gemini-1.5-flash} and \texttt{Gemini-1.5-flash-lite} (specifically, the \texttt{001} series), which achieved similarly competitive performance, in \Cref{sec:base_model_comparisons}.

\subsubsection{Baselines.} 
\begin{figure}[h]
    \centering
    \includegraphics[width=0.45\textwidth]{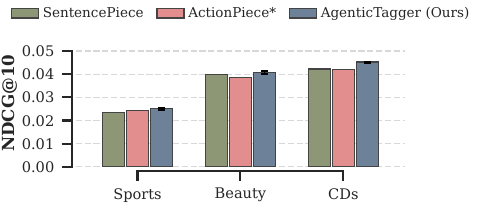}
    \caption{Comparison against feature-crossing-based methods in NDCG@10. For ActionPiece, we compare against its ablated variant \textit{without} inference time model ensembling, to isolate the effect on item-content understanding.}
    \label{fig:apc_spm_comparison}
\end{figure}

In our main experiment (\Cref{tab:performance_comparison_public}), we compare our method with several families of models: models with item IDs only, such as SASRec~\cite{Kang2018SelfAttentiveSR} and BERT4Rec~\cite{fei2019bert4rec}; models enhanced by embedded features: FDSA~\cite{zhang2019FDSA}, $\text{S}^3$-Rec~\cite{zhou2020s3rec}; and models with discrete features (i.e., semantic IDs) derived from semantic information or collaborative information: VQ-Rec~\cite{hou2023vqrec}, TIGER~\cite{rajput2023recommender}, HSTU~\cite{zhao2024hstu}, and LLM-Indexer~\cite{jin2024llmindexer}.
We additionally compare our method with models that enhance raw semantic IDs via feature crossing, namely SPM-CID~\cite{Anima2024Better} and ActionPiece~\cite{hou2025actionpiece} in~\Cref{fig:apc_spm_comparison}.

\subsubsection{Main Results.} We summarize our results on the Amazon Reviews benchmarks in~\Cref{tab:performance_comparison_public}.
Overall, models with sets of discrete item features outperform other competing methods that make use of single-embedding-based item features, while~\ourmethodauto helps achieve the best overall performance compared to various baselines, demonstrating its effectiveness.

Interestingly, the performance improvements are more significant in the Sports and CDs domains, with 8.9\%-11.6\% improvements over the closest competing model in NDCG@10, whereas the improvement is moderate (2\%) in the Beauty domain. 
We hypothesize that this could be because \ourmethodauto yields the most benefit when it is able to construct its vocabulary from a larger item corpus, where the natural language feedback aggregated via the $\text{Distill}$ operation provides more stable signals. 
We further discuss this hypothesis in~\Cref{subsubsec:scaling_to_larger_dataset}, where we show that \ourmethodauto benefits from larger datasets with diverse items, even when doing so at the cost of pruning the vocabulary-building process.

\subsubsection{Comparisons with ID Crossing.}
In light of recent advances in demonstrating the benefit of incorporating dataset collaborative filtering statistics directly into the tokenization process via feature crossing~\cite{hou2025actionpiece, Anima2024Better}, we also compare our method with two recent tokenization methods that merge frequently co-occurring semantic IDs into new IDs based on observed user behavior sequences, SentencePiece~\cite{Anima2024Better} and ActionPiece~\cite{hou2025actionpiece}. 
To be specific, these methods represent an alternative path towards better features for recommender systems: automated feature crossing based on user behavior.
As shown in~\Cref{fig:apc_spm_comparison}, \ourmethodauto remains a strong tokenization method, demonstrating the effectiveness of leveraging LLMs' content understanding abilities in tokenization in addition to collaborative filtering signals.
Meanwhile, we note that \ourmethodauto is orthogonal to these crossing-based techniques, and we leave the crossing of \ourmethodauto features to future study.

\subsection{Ranking}
\label{subsec:ranking}

\begin{table}[h]
\centering
\caption{Ranking performance comparison between the Item ID baseline and \ourmethod.}
\label{tab:ranking_results_full}
\begin{tabular}{lcc}
\toprule
\textbf{Metric} & \textbf{Item ID (Baseline)} & \textbf{\ourmethod (Ours)} \\
\midrule
R@5  & 0.2655 $\pm$ 0.0037 & \textbf{0.2823 $\pm$ 0.0038} \\
N@5    & \textbf{0.1966 $\pm$ 0.0029} & 0.1870 $\pm$ 0.0027 \\
\midrule
R@10 & 0.3456 $\pm$ 0.0040 & \textbf{0.4300 $\pm$ 0.0041} \\
N@10   & 0.2224 $\pm$ 0.0029 & \textbf{0.2346 $\pm$ 0.0027} \\
\midrule
R@20 & 0.4427 $\pm$ 0.0042 & \textbf{0.6082 $\pm$ 0.0041} \\
N@20   & 0.2470 $\pm$ 0.0029 & \textbf{0.2796 $\pm$ 0.0025} \\
\midrule
R@50 & 0.5761 $\pm$ 0.0041 & \textbf{0.8514 $\pm$ 0.0030} \\
N@50   & 0.2733 $\pm$ 0.0027 & \textbf{0.3280 $\pm$ 0.0022} \\
\bottomrule
\end{tabular}
\end{table}

\subsubsection{Dataset, Baseline, and Evaluation.} Besides generative recommendation, another frequent use case for structured item features is ranking~\cite{Anima2024Better, zhu2021bars}. 
To this end, we evaluate the effect of \ourmethod features on a private dataset collected from a news-feed serving platform over a random subset of 47,358 users across an 8-day period. This results in 67,097 items and 206,345 interactions, where raw item content and user interactions are fully accessible.
Following prior work in the literature~\cite{he2017neuralcollaborativefiltering, Kang2018SelfAttentiveSR}, we randomly sample 100 negative items per user to reduce computational overhead, and report performance with Recall@$K$ and NDCG@$K$, where $K \in \{5, 10, 20, 50\}$, similar to our experiments on public benchmarks.
The backbone recommendation model is a neural feature-aware model aligned with the deployed model in the platform, where we use either one-hot item IDs (i.e., one embedding per item) or \ourmethod descriptors as IDs during training. 
For good scalability, we adopt a 3-layer setting for \ourmethodauto, and build the vocabulary using 30,000 randomly subsampled items, resulting in 3 descriptors per item. 
We break item collisions using an extra digit of a collision-resolver token following standard practice in the semantic ID literature~\cite{rajput2023recommender, hou2025rpg, he2025plumadaptingpretrainedlanguage, Anima2024Better}.

\subsubsection{Overall Performance.} We report our results in~\Cref{tab:ranking_results_full}; as shown, \ourmethod brings significant improvements to ranking accuracy compared to raw item IDs, showcasing the effectiveness of \ourmethod features.
On the other hand, an interesting observation is that the item ID baseline shows slightly stronger performance for NDCG@5. 
We hypothesize that this is because when the dataset is sufficiently large for good convergence, as in our case, memorization of item-specific collaborative filtering signals via unique per-item embeddings helps fine-grained ranking among top candidates. 
This is aligned with prior findings in the literature~\cite{kang2021learning}, where compressing one-hot item representations into hashed embedding tables sometimes requires (albeit marginally) a performance trade-off.
We discuss further verification of this idea in the appendix, where we show that in smaller datasets where interactions are sparser, \ourmethod-generated features exhibit larger improvements compared to one-hot item representations (\Cref{subsec:ranking_performance_smaller_dataset}).

\subsubsection{Cardinality.} As discussed in the previous paragraph, \ourmethodauto achieves strong performance with a much lower-cardinality vocabulary, utilizing only 2.7k unique descriptors compared to the full one-hot item table of over 67k item embeddings, achieving effective vocabulary compression. 
This demonstrates that \ourmethod effectively captures the distinctive characteristics of each item within its automatically inferred descriptor vocabulary, validating the effectiveness of our proposed framework in enabling LLMs to understand and capture frequent patterns in the item corpus.

\subsection{Interpretability and Controllability}
\label{subsec:interpretability_and_controllability}

\begin{table}[h] 
    \centering 
    \setlength{\tabcolsep}{4pt} 
    \small 
    \caption{Performance improvements after single-turn critique on 5000 samples.}
    \label{tab:interaction_performance} 
    \begin{tabular}{llccc}
        \toprule
        \textbf{Dataset} & \textbf{Metric} & \textbf{Vanilla} & \textbf{+Critique} & \textbf{Improve.} \\
        \midrule
        \multirow{4}{*}{\textbf{Sports}} 
         & R@5  & 0.0318 & 0.0428 & +34.59\% \\
         & N@5  & 0.0206 & 0.0293 & +42.09\% \\
         & R@10 & 0.0500 & 0.0670 & +34.00\% \\
         & N@10 & 0.0264 & 0.0371 & +40.31\% \\
        \midrule
        \multirow{4}{*}{\textbf{Beauty}} 
         & R@5  & 0.0504 & 0.0864 & +71.43\% \\
         & N@5  & 0.0350 & 0.0602 & +72.17\% \\
         & R@10 & 0.0738 & 0.1284 & +73.98\% \\
         & N@10 & 0.0425 & 0.0738 & +73.65\% \\
        \midrule
        \multirow{4}{*}{\textbf{CDs}} 
         & R@5  & 0.0546 & 0.0666 & +21.98\% \\
         & N@5  & 0.0374 & 0.0445 & +19.21\% \\
         & R@10 & 0.0816 & 0.0990 & +21.32\% \\
         & N@10 & 0.0460 & 0.0549 & +19.42\% \\
        \bottomrule
    \end{tabular}
\end{table}

As discussed earlier, a core advantage of LLM-generated features is their explainability, which naturally suits user-facing use cases, such as situations where one would like to control or constrain the matching mechanism of a recommender system. 
In this section, we discuss the value \ourmethodauto features add to the interpretability and controllability of a recommender system via a single-turn critique setting, where a user gives feedback to the recommender system, providing information on the item they would like to interact with next.
In particular, this setting is a fundamental feature in many conversational recommender systems~\cite{Raymond2018Towards, wang2022towards, He2023Large, zhu-etal-2025-llm-based}, and commonly manifests in production recommender systems as feedback with keyphrase-based rationales.

Inspired by recent work on using LLMs for user simulation in recommender systems~\cite{bougie-watanabe-2025-simuser, park2025LLMasUserSim, Zhang2025LLMPoweredUserSim, wang2024userbehavsim} and conversational recommendation~\cite{wang-etal-2023-rethinking-evaluation, zhu25allmbased}, we make use of an LLM as a user simulator to evaluate whether~\ourmethodauto can enable effective user control over recommender system behavior.
Specifically, we adopt a communication game in an information bottleneck setting, where an LLM (\texttt{gemini-2.5-flash} in practice) observes the descriptions of target items in its prompt, and is asked to select all potentially relevant top-level (i.e., level 1) features in the vocabulary, which generally contains a small number of keywords (e.g., 16 for the CDs and Vinyl domain, the largest public benchmark we evaluate on).
We then generate the next item from a pre-trained generative recommender backbone using beam search, following standard practice, except we constrain its search space to be within the allowed level 1 features provided by the user simulator.
While this interaction paradigm essentially ``leaks'' ground-truth information to the recommender, it verifies that (1) \ourmethodauto produces a vocabulary that precisely identifies a meaningful subset of items, and (2) the produced vocabulary aligns well with the semantic understanding of LLMs, which are known to align well with human perception~\cite{chang-bergen-2024-language}, demonstrating the interpretability and controllability potential of \ourmethodauto features.

As shown in~\Cref{tab:interaction_performance}, this constrained recommendation setting increases model performance significantly across datasets, showcasing that \ourmethod features effectively communicate the nuances of items in the dataset and, importantly, in an interpretable manner, such that the behavior of the corresponding recommender system using \ourmethodauto features can be straightforwardly controlled by natural-language-based critiques.

\section{Ablations and Discussions}
\label{subsec:ablations}

\subsection{Effect of Multi-agent Self-refinement}

\begin{figure}[t]
    \centering
    \includegraphics[width=\columnwidth, trim={0pt 0pt 0pt 0pt}, clip]{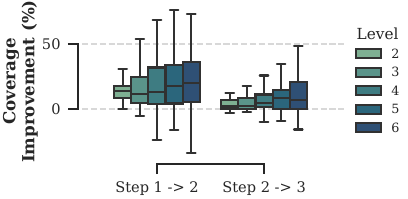}
    \caption{
        Distribution of coverage changes between optimization steps across layers. We report level 2 to level 6, which have enough trials to obtain meaningful results for analysis.
    }
    \label{fig:sports_iter_refine_ablation}
\end{figure}

In this ablation, we examine the effect of the iterative refinement loop in~\ourmethodauto on improving item coverage.
To verify its effect, we plot the average percentage differences in item coverage (i.e., items that are assigned at least one feature among the current vocabulary) for each clustering job in~\Cref{fig:sports_iter_refine_ablation}.
As shown, the overall trend is that iterative refinement helps improve item coverage, leading to a reduced number of items that cannot be described by the vocabulary as the number of optimization steps grows.
While we report the difference between steps 1 and 2 on the Sports dataset, the trend remains consistent across datasets and optimization steps. 

On the other hand, we observe that iterative optimization tends to bring less steady coverage improvements to later-layer (i.e., finer-grained) \ourmethodauto features.
To this end, we hypothesize that to effectively improve item coverage, the LLM must be able to understand increasingly nuanced item-to-item similarities as the hierarchy depth increases; intuitively, it is easier to separate high-level differences, such as Rock music and Blues music, than specific subgenres of Rock music.

\subsection{Scaling \ourmethodauto to Larger Datasets}
\label{subsubsec:scaling_to_larger_dataset}

In \ourmethodauto, a key decision to make when scaling to larger datasets is the balance between controlling the maximum branching factor $b$, i.e., how many features an item can be associated with at each granularity level, and how much data to subsample for generating the vocabulary using~\Cref{alg:hierarchical}.
While these two techniques are not mutually exclusive, it is nevertheless helpful to understand which tends to yield better performance for a large dataset. 

To investigate this trade-off, we compare the quality of~\ourmethodauto features generated from a random subset of 20k items with that of the original scaling choice we made on the CDs and Vinyl domain, where we constrained the branching factor to 1.
That is, we only consider one randomly selected associated feature at each feature level for an item when generating finer-grained features in the next level.
We found that, surprisingly, constraining the branching factor yields significantly superior performance compared to subsampling, where the performance drops to a level comparable to P5-CID. 
We hypothesize that constructing the vocabulary from more items helps \ourmethodauto assign valid features to long-tail item groups, which are crucial for capturing niche user interests.

\subsection{Comparison with Free-form Generation in Generative Recommendation}
\label{subsec:free_form_tag_as_semid}

As discussed earlier, it is difficult to directly use free-form generated features for generative recommendation, as this method assigns sets of features to items with no constraints on vocabulary choice.
To illustrate this point, we discuss an earlier attempt aimed at directly using free-form generated features as semantic IDs in the Sports domain.

We first prompt an LLM to generate free-form descriptors and treat all ever-generated descriptors as part of the valid vocabulary.
We then train a dual-encoder model that predicts which descriptors are generated for each item among the possible descriptor vocabulary (i.e., every descriptor that has been generated by the LLM), following prior work in LLM-based document topic mining~\cite{xie-etal-2025-latent}, where positive signals come from LLM generations. 
This process ensures that the potential match between each descriptor-item pair is sufficiently explored, even when the LLM did not generate a particular descriptor initially. 
Finally, we perform retrieval-augmented descriptor generation with another pass of prompting, except that the LLM this time is asked to select all relevant descriptors from the top 30 suggested descriptors predicted by the aforementioned dual-encoder model. This process yields a maximum of 27, a minimum of 0, and an average of 12.25 descriptors per item.

Due to the large number of descriptors per item, we then prune the descriptors assigned to each item via two methods. (1) Frequency: we split the tags by their occurrence across the corpus into four equal bins, ignoring tags with frequencies smaller than 10 or larger than 2,000, since these are overly rare or common. Each item's descriptors are then selected by taking the most frequent tag from each popularity bin, where available, creating a coarse-to-fine tag sequence similar to semantic IDs. (2) K-means: we select 5,000 centroids by running K-means clustering over the text embeddings of the descriptor vocabulary, and swap the descriptors with their assigned centroid IDs, reducing the number of descriptors per item.

However, this framework results in poor overall performance: $\text{NDCG}@10$ for frequency-based pruning is 0.185, and for K-means-based pruning, it is 0.184, neither of which is comparable to modern semantic ID performance. 
This demonstrates, in contrast, the effectiveness of \ourmethodauto in generating hierarchically structured, semantic-ID-like descriptors.

\begin{table}[t]
\centering
\caption{Comparison on \ourmethod performance with different LLMs on generative recommendation.}
\label{tab:ablation_backbones_compact}
\small
\setlength{\tabcolsep}{3pt}
\begin{tabular}{@{}l cc cc cc@{}}
\toprule
\multirow{2}{*}{\textbf{Backbone}} & \multicolumn{2}{c}{\textbf{Sports}} & \multicolumn{2}{c}{\textbf{Beauty}} & \multicolumn{2}{c}{\textbf{CDs}} \\
\cmidrule(lr){2-3} \cmidrule(lr){4-5} \cmidrule(lr){6-7}
 & N@5 & N@10 & N@5 & N@10 & N@5 & N@10 \\
\midrule
Gemini-1.5 & 0.0184 & 0.0237 & 0.0319 & 0.0397 & 0.0349 & 0.0439 \\
Gemini-2.5 & \textbf{0.0194} & \textbf{0.0251} & \textbf{0.0332} & \textbf{0.0408} & \textbf{0.0358} & \textbf{0.0453} \\
\bottomrule
\end{tabular}
\end{table}

\subsection{Performance with Different LLMs}
\label{sec:base_model_comparisons}

In our generative recommendation experiments on public benchmarks, we opted to use the \texttt{gemini-2.5} family of models, as they were released closest to the time the experiments were conducted. 
However, to investigate if our framework is robust to the choice of base models, we also experimented with \texttt{gemini-1.5-flash} and \texttt{gemini-1.5-pro} as annotator and architect LLMs, respectively.
In our experiments, we find both models yield competitive performance, while \texttt{gemini-2.5} maintains a slight advantage across datasets, likely due to its stronger overall abilities in text generation and understanding.

\subsection{Case Study}

\begin{table}[t]
\centering
\small 
\caption{Case Study: \ourmethod and free-form generation as features. Item content rewritten to preserve privacy.}
\label{tab:case_study}
\begin{tabularx}{\columnwidth}{@{}p{0.3\columnwidth} X X @{}}
\toprule
\textbf{Item Context} & \textbf{Free-form} & \textbf{\ourmethod} \\ 
\midrule
\textbf{Ground Truth} & Skipper Limited, & \textbf{L0:} Business \& Finance \\
\textit{Transformer stock,} & Order Book Growth, & \textbf{L1:} Stock Market \\
\textit{820\% 3-yr returns} & Power Transmission. & \textbf{L2:} Individual Performance \\
\midrule
\textbf{User History} & & \textbf{L0:} Business \& Finance \\
(1) \textit{Defence bidder, 7k\% 5-yr returns} & (1) Defence PSU, DRDO Bidding. & \textbf{L1:} Stock Market \\
(2) \textit{Industrial PSU, 200\% 5-mo returns} & (2) PSU Stocks, Market Returns. & \textbf{L2:} Individual Performance, Order Book Analysis \\
\bottomrule
\end{tabularx}
\vspace{-1em}
\end{table}

Finally, to understand the improvement brought by \ourmethodauto compared to free-form generation, we took an example from our private dataset to analyze the effect of features from both frameworks on connecting the user's recent interests to the ground-truth item.
As shown, the free-form-generated features, while of generally high coherence, failed to generalize across items due to random surface phrase choices; in contrast, \ourmethodauto successfully identified the user's broad interest in the stock market and detailed interest in stock performance. 
While in both cases LLMs are effective at extracting general user interest in stocks, \ourmethod is the only method that can organically connect user history to the ground-truth interacted item, demonstrating the value of vocabulary-constrained, structured descriptor generation.

\section{Conclusion}

In this work, we introduce \ourmethod, a general framework for prompting LLMs to generate natural language item-descriptor representations for recommender systems. We showcase the wide applicability of \ourmethod across various applications, including generative retrieval, ranking, and controllability-oriented, critique-based recommendation.
We also discuss the properties of \ourmethod, such as its complexity and feature cardinality. In the future, we plan to investigate methods for improving effectiveness by introducing \ourmethod implementations that consider collaborative filtering signals during the feature generation phase, following the recent success of automated feature crossing.

\bibliographystyle{ACM-Reference-Format}
\bibliography{references}


\begin{thebibliography}{84}


\ifx \showCODEN    \undefined \def \showCODEN     #1{\unskip}     \fi
\ifx \showISBNx    \undefined \def \showISBNx     #1{\unskip}     \fi
\ifx \showISBNxiii \undefined \def \showISBNxiii  #1{\unskip}     \fi
\ifx \showISSN     \undefined \def \showISSN      #1{\unskip}     \fi
\ifx \showLCCN     \undefined \def \showLCCN      #1{\unskip}     \fi
\ifx \shownote     \undefined \def \shownote      #1{#1}          \fi
\ifx \showarticletitle \undefined \def \showarticletitle #1{#1}   \fi
\ifx \showURL      \undefined \def \showURL       {\relax}        \fi
\providecommand\bibfield[2]{#2}
\providecommand\bibinfo[2]{#2}
\providecommand\natexlab[1]{#1}
\providecommand\showeprint[2][]{arXiv:#2}

\bibitem[Bougie and Watanabe(2025)]%
        {bougie-watanabe-2025-simuser}
\bibfield{author}{\bibinfo{person}{Nicolas Bougie} {and} \bibinfo{person}{Narimawa Watanabe}.} \bibinfo{year}{2025}\natexlab{}.
\newblock \showarticletitle{{S}im{USER}: Simulating User Behavior with Large Language Models for Recommender System Evaluation}. In \bibinfo{booktitle}{\emph{Proceedings of the 63rd Annual Meeting of the Association for Computational Linguistics (Volume 6: Industry Track)}}, \bibfield{editor}{\bibinfo{person}{Georg Rehm} {and} \bibinfo{person}{Yunyao Li}} (Eds.). \bibinfo{publisher}{Association for Computational Linguistics}, \bibinfo{address}{Vienna, Austria}, \bibinfo{pages}{43--60}.
\newblock
\showISBNx{979-8-89176-288-6}
\href{https://doi.org/10.18653/v1/2025.acl-industry.5}{doi:\nolinkurl{10.18653/v1/2025.acl-industry.5}}


\bibitem[Chang and Bergen(2024)]%
        {chang-bergen-2024-language}
\bibfield{author}{\bibinfo{person}{Tyler~A. Chang} {and} \bibinfo{person}{Benjamin~K. Bergen}.} \bibinfo{year}{2024}\natexlab{}.
\newblock \showarticletitle{Language Model Behavior: A Comprehensive Survey}.
\newblock \bibinfo{journal}{\emph{Computational Linguistics}} \bibinfo{volume}{50}, \bibinfo{number}{1} (\bibinfo{date}{March} \bibinfo{year}{2024}), \bibinfo{pages}{293--350}.
\newblock
\href{https://doi.org/10.1162/coli_a_00492}{doi:\nolinkurl{10.1162/coli_a_00492}}


\bibitem[Cheng et~al\mbox{.}(2016)]%
        {cheng2016widedeep}
\bibfield{author}{\bibinfo{person}{Heng-Tze Cheng}, \bibinfo{person}{Levent Koc}, \bibinfo{person}{Jeremiah Harmsen}, \bibinfo{person}{Tal Shaked}, \bibinfo{person}{Tushar Chandra}, \bibinfo{person}{Hrishi Aradhye}, \bibinfo{person}{Glen Anderson}, \bibinfo{person}{Greg Corrado}, \bibinfo{person}{Wei Chai}, \bibinfo{person}{Mustafa Ispir}, \bibinfo{person}{Rohan Anil}, \bibinfo{person}{Zakaria Haque}, \bibinfo{person}{Lichan Hong}, \bibinfo{person}{Vihan Jain}, \bibinfo{person}{Xiaobing Liu}, {and} \bibinfo{person}{Hemal Shah}.} \bibinfo{year}{2016}\natexlab{}.
\newblock \showarticletitle{Wide \& Deep Learning for Recommender Systems}. In \bibinfo{booktitle}{\emph{Proceedings of the 1st Workshop on Deep Learning for Recommender Systems}} (Boston, MA, USA) \emph{(\bibinfo{series}{DLRS 2016})}. \bibinfo{publisher}{Association for Computing Machinery}, \bibinfo{address}{New York, NY, USA}, \bibinfo{pages}{7–10}.
\newblock
\showISBNx{9781450347952}
\href{https://doi.org/10.1145/2988450.2988454}{doi:\nolinkurl{10.1145/2988450.2988454}}


\bibitem[Coleman et~al\mbox{.}(2023)]%
        {coleman2023unified}
\bibfield{author}{\bibinfo{person}{Benjamin Coleman}, \bibinfo{person}{Wang-Cheng Kang}, \bibinfo{person}{Matthew Fahrbach}, \bibinfo{person}{Ruoxi Wang}, \bibinfo{person}{Lichan Hong}, \bibinfo{person}{Ed~H. Chi}, {and} \bibinfo{person}{Derek~Zhiyuan Cheng}.} \bibinfo{year}{2023}\natexlab{}.
\newblock \showarticletitle{Unified Embedding: Battle-Tested Feature Representations for Web-Scale {ML} Systems}. In \bibinfo{booktitle}{\emph{Thirty-seventh Conference on Neural Information Processing Systems}}.
\newblock
\urldef\tempurl%
\url{https://openreview.net/forum?id=hJzEoQHfCe}
\showURL{%
\tempurl}


\bibitem[Friedman et~al\mbox{.}(2023)]%
        {friedman2023leveraginglargelanguagemodels}
\bibfield{author}{\bibinfo{person}{Luke Friedman}, \bibinfo{person}{Sameer Ahuja}, \bibinfo{person}{David Allen}, \bibinfo{person}{Zhenning Tan}, \bibinfo{person}{Hakim Sidahmed}, \bibinfo{person}{Changbo Long}, \bibinfo{person}{Jun Xie}, \bibinfo{person}{Gabriel Schubiner}, \bibinfo{person}{Ajay Patel}, \bibinfo{person}{Harsh Lara}, \bibinfo{person}{Brian Chu}, \bibinfo{person}{Zexi Chen}, {and} \bibinfo{person}{Manoj Tiwari}.} \bibinfo{year}{2023}\natexlab{}.
\newblock \bibinfo{title}{Leveraging Large Language Models in Conversational Recommender Systems}.
\newblock
\showeprint[arxiv]{2305.07961}~[cs.IR]
\urldef\tempurl%
\url{https://arxiv.org/abs/2305.07961}
\showURL{%
\tempurl}


\bibitem[Gao et~al\mbox{.}(2022)]%
        {Gao2022KuaiRec}
\bibfield{author}{\bibinfo{person}{Chongming Gao}, \bibinfo{person}{Shijun Li}, \bibinfo{person}{Wenqiang Lei}, \bibinfo{person}{Jiawei Chen}, \bibinfo{person}{Biao Li}, \bibinfo{person}{Peng Jiang}, \bibinfo{person}{Xiangnan He}, \bibinfo{person}{Jiaxin Mao}, {and} \bibinfo{person}{Tat-Seng Chua}.} \bibinfo{year}{2022}\natexlab{}.
\newblock \showarticletitle{KuaiRec: A Fully-observed Dataset and Insights for Evaluating Recommender Systems}. In \bibinfo{booktitle}{\emph{Proceedings of the 31st ACM International Conference on Information \& Knowledge Management}} (Atlanta, GA, USA) \emph{(\bibinfo{series}{CIKM '22})}. \bibinfo{publisher}{Association for Computing Machinery}, \bibinfo{address}{New York, NY, USA}, \bibinfo{pages}{540–550}.
\newblock
\showISBNx{9781450392365}
\href{https://doi.org/10.1145/3511808.3557220}{doi:\nolinkurl{10.1145/3511808.3557220}}


\bibitem[Ge et~al\mbox{.}(2014)]%
        {Ge2014OptimizedPQ}
\bibfield{author}{\bibinfo{person}{Tiezheng Ge}, \bibinfo{person}{Kaiming He}, \bibinfo{person}{Qifa Ke}, {and} \bibinfo{person}{Jian Sun}.} \bibinfo{year}{2014}\natexlab{}.
\newblock \showarticletitle{Optimized Product Quantization}.
\newblock \bibinfo{journal}{\emph{IEEE Transactions on Pattern Analysis and Machine Intelligence}}  \bibinfo{volume}{36} (\bibinfo{year}{2014}), \bibinfo{pages}{744--755}.
\newblock
\urldef\tempurl%
\url{https://api.semanticscholar.org/CorpusID:6033212}
\showURL{%
\tempurl}


\bibitem[{Google Ads Help}(2026)]%
        {google2026aboutkeywords}
\bibfield{author}{\bibinfo{person}{{Google Ads Help}}.} \bibinfo{year}{2026}\natexlab{}.
\newblock \bibinfo{title}{About keywords in Search Network campaigns}.
\newblock \bibinfo{howpublished}{\url{https://support.google.com/google-ads/answer/1704371}}.
\newblock
\newblock
\shownote{Accessed: 2026-01-28}.


\bibitem[Harper and Konstan(2015)]%
        {harper2016movielens}
\bibfield{author}{\bibinfo{person}{F.~Maxwell Harper} {and} \bibinfo{person}{Joseph~A. Konstan}.} \bibinfo{year}{2015}\natexlab{}.
\newblock \showarticletitle{The MovieLens Datasets: History and Context}.
\newblock \bibinfo{journal}{\emph{ACM Trans. Interact. Intell. Syst.}} \bibinfo{volume}{5}, \bibinfo{number}{4}, Article \bibinfo{articleno}{19} (\bibinfo{date}{Dec.} \bibinfo{year}{2015}), \bibinfo{numpages}{19}~pages.
\newblock
\showISSN{2160-6455}
\href{https://doi.org/10.1145/2827872}{doi:\nolinkurl{10.1145/2827872}}


\bibitem[He et~al\mbox{.}(2025)]%
        {he2025plumadaptingpretrainedlanguage}
\bibfield{author}{\bibinfo{person}{Ruining He}, \bibinfo{person}{Lukasz Heldt}, \bibinfo{person}{Lichan Hong}, \bibinfo{person}{Raghunandan Keshavan}, \bibinfo{person}{Shifan Mao}, \bibinfo{person}{Nikhil Mehta}, \bibinfo{person}{Zhengyang Su}, \bibinfo{person}{Alicia Tsai}, \bibinfo{person}{Yueqi Wang}, \bibinfo{person}{Shao-Chuan Wang}, \bibinfo{person}{Xinyang Yi}, \bibinfo{person}{Lexi Baugher}, \bibinfo{person}{Baykal Cakici}, \bibinfo{person}{Ed Chi}, \bibinfo{person}{Cristos Goodrow}, \bibinfo{person}{Ningren Han}, \bibinfo{person}{He Ma}, \bibinfo{person}{Romer Rosales}, \bibinfo{person}{Abby~Van Soest}, \bibinfo{person}{Devansh Tandon}, \bibinfo{person}{Su-Lin Wu}, \bibinfo{person}{Weilong Yang}, {and} \bibinfo{person}{Yilin Zheng}.} \bibinfo{year}{2025}\natexlab{}.
\newblock \bibinfo{title}{PLUM: Adapting Pre-trained Language Models for Industrial-scale Generative Recommendations}.
\newblock
\showeprint[arxiv]{2510.07784}~[cs.IR]
\urldef\tempurl%
\url{https://arxiv.org/abs/2510.07784}
\showURL{%
\tempurl}


\bibitem[He et~al\mbox{.}(2017)]%
        {he2017neuralcollaborativefiltering}
\bibfield{author}{\bibinfo{person}{Xiangnan He}, \bibinfo{person}{Lizi Liao}, \bibinfo{person}{Hanwang Zhang}, \bibinfo{person}{Liqiang Nie}, \bibinfo{person}{Xia Hu}, {and} \bibinfo{person}{Tat-Seng Chua}.} \bibinfo{year}{2017}\natexlab{}.
\newblock \bibinfo{title}{Neural Collaborative Filtering}.
\newblock
\showeprint[arxiv]{1708.05031}~[cs.IR]
\urldef\tempurl%
\url{https://arxiv.org/abs/1708.05031}
\showURL{%
\tempurl}


\bibitem[He et~al\mbox{.}(2023)]%
        {He2023Large}
\bibfield{author}{\bibinfo{person}{Zhankui He}, \bibinfo{person}{Zhouhang Xie}, \bibinfo{person}{Rahul Jha}, \bibinfo{person}{Harald Steck}, \bibinfo{person}{Dawen Liang}, \bibinfo{person}{Yesu Feng}, \bibinfo{person}{Bodhisattwa~Prasad Majumder}, \bibinfo{person}{Nathan Kallus}, {and} \bibinfo{person}{Julian Mcauley}.} \bibinfo{year}{2023}\natexlab{}.
\newblock \showarticletitle{Large Language Models as Zero-Shot Conversational Recommenders}. In \bibinfo{booktitle}{\emph{Proceedings of the 32nd ACM International Conference on Information and Knowledge Management}} (Birmingham, United Kingdom) \emph{(\bibinfo{series}{CIKM '23})}. \bibinfo{publisher}{Association for Computing Machinery}, \bibinfo{address}{New York, NY, USA}, \bibinfo{pages}{720–730}.
\newblock
\showISBNx{9798400701245}
\href{https://doi.org/10.1145/3583780.3614949}{doi:\nolinkurl{10.1145/3583780.3614949}}


\bibitem[Hou et~al\mbox{.}(2023)]%
        {hou2023vqrec}
\bibfield{author}{\bibinfo{person}{Yupeng Hou}, \bibinfo{person}{Zhankui He}, \bibinfo{person}{Julian McAuley}, {and} \bibinfo{person}{Wayne~Xin Zhao}.} \bibinfo{year}{2023}\natexlab{}.
\newblock \showarticletitle{Learning Vector-Quantized Item Representation for Transferable Sequential Recommenders}. In \bibinfo{booktitle}{\emph{{TheWebConf}}}.
\newblock


\bibitem[Hou et~al\mbox{.}(2025a)]%
        {hou2025rpg}
\bibfield{author}{\bibinfo{person}{Yupeng Hou}, \bibinfo{person}{Jiacheng Li}, \bibinfo{person}{Ashley Shin}, \bibinfo{person}{Jinsung Jeon}, \bibinfo{person}{Abhishek Santhanam}, \bibinfo{person}{Wei Shao}, \bibinfo{person}{Kaveh Hassani}, \bibinfo{person}{Ning Yao}, {and} \bibinfo{person}{Julian McAuley}.} \bibinfo{year}{2025}\natexlab{a}.
\newblock \showarticletitle{Generating Long Semantic IDs in Parallel for Recommendation}. In \bibinfo{booktitle}{\emph{{KDD}}}.
\newblock


\bibitem[Hou et~al\mbox{.}(2025b)]%
        {hou2025actionpiece}
\bibfield{author}{\bibinfo{person}{Yupeng Hou}, \bibinfo{person}{Jianmo Ni}, \bibinfo{person}{Zhankui He}, \bibinfo{person}{Noveen Sachdeva}, \bibinfo{person}{Wang-Cheng Kang}, \bibinfo{person}{Ed~H. Chi}, \bibinfo{person}{Julian McAuley}, {and} \bibinfo{person}{Derek~Zhiyuan Cheng}.} \bibinfo{year}{2025}\natexlab{b}.
\newblock \showarticletitle{ActionPiece: Contextually Tokenizing Action Sequences for Generative Recommendation}. In \bibinfo{booktitle}{\emph{Forty-second International Conference on Machine Learning}}.
\newblock
\urldef\tempurl%
\url{https://openreview.net/forum?id=h2oNQOzbc5}
\showURL{%
\tempurl}


\bibitem[Huang et~al\mbox{.}(2025)]%
        {huang2025recaiagent}
\bibfield{author}{\bibinfo{person}{Xu Huang}, \bibinfo{person}{Jianxun Lian}, \bibinfo{person}{Yuxuan Lei}, \bibinfo{person}{Jing Yao}, \bibinfo{person}{Defu Lian}, {and} \bibinfo{person}{Xing Xie}.} \bibinfo{year}{2025}\natexlab{}.
\newblock \showarticletitle{Recommender AI Agent: Integrating Large Language Models for Interactive Recommendations}.
\newblock \bibinfo{journal}{\emph{ACM Trans. Inf. Syst.}} \bibinfo{volume}{43}, \bibinfo{number}{4}, Article \bibinfo{articleno}{96} (\bibinfo{date}{June} \bibinfo{year}{2025}), \bibinfo{numpages}{33}~pages.
\newblock
\showISSN{1046-8188}
\href{https://doi.org/10.1145/3731446}{doi:\nolinkurl{10.1145/3731446}}


\bibitem[Jalan et~al\mbox{.}(2024)]%
        {jalan2024llmbrec}
\bibfield{author}{\bibinfo{person}{Raksha Jalan}, \bibinfo{person}{Tushar Prakash}, {and} \bibinfo{person}{Niranjan Pedanekar}.} \bibinfo{year}{2024}\natexlab{}.
\newblock \showarticletitle{{LLM}-{BR}ec: Personalizing Session-based Social Recommendation with {LLM}-{BERT} Fusion Framework}. In \bibinfo{booktitle}{\emph{The Second Workshop on Generative Information Retrieval}}.
\newblock
\urldef\tempurl%
\url{https://openreview.net/forum?id=gwHVlTNKsG}
\showURL{%
\tempurl}


\bibitem[Jegou et~al\mbox{.}(2011)]%
        {Jegou2011ProductQ}
\bibfield{author}{\bibinfo{person}{Herve Jegou}, \bibinfo{person}{Matthijs Douze}, {and} \bibinfo{person}{Cordelia Schmid}.} \bibinfo{year}{2011}\natexlab{}.
\newblock \showarticletitle{Product Quantization for Nearest Neighbor Search}.
\newblock \bibinfo{journal}{\emph{IEEE Trans. Pattern Anal. Mach. Intell.}} \bibinfo{volume}{33}, \bibinfo{number}{1} (\bibinfo{date}{Jan.} \bibinfo{year}{2011}), \bibinfo{pages}{117–128}.
\newblock
\showISSN{0162-8828}
\href{https://doi.org/10.1109/TPAMI.2010.57}{doi:\nolinkurl{10.1109/TPAMI.2010.57}}


\bibitem[Jiang et~al\mbox{.}(2025)]%
        {jiang-etal-2025-reclm}
\bibfield{author}{\bibinfo{person}{Yangqin Jiang}, \bibinfo{person}{Yuhao Yang}, \bibinfo{person}{Lianghao Xia}, \bibinfo{person}{Da Luo}, \bibinfo{person}{Kangyi Lin}, {and} \bibinfo{person}{Chao Huang}.} \bibinfo{year}{2025}\natexlab{}.
\newblock \showarticletitle{{R}ec{LM}: Recommendation Instruction Tuning}. In \bibinfo{booktitle}{\emph{Proceedings of the 63rd Annual Meeting of the Association for Computational Linguistics (Volume 1: Long Papers)}}, \bibfield{editor}{\bibinfo{person}{Wanxiang Che}, \bibinfo{person}{Joyce Nabende}, \bibinfo{person}{Ekaterina Shutova}, {and} \bibinfo{person}{Mohammad~Taher Pilehvar}} (Eds.). \bibinfo{publisher}{Association for Computational Linguistics}, \bibinfo{address}{Vienna, Austria}, \bibinfo{pages}{15443--15459}.
\newblock
\showISBNx{979-8-89176-251-0}
\href{https://doi.org/10.18653/v1/2025.acl-long.751}{doi:\nolinkurl{10.18653/v1/2025.acl-long.751}}


\bibitem[Jin et~al\mbox{.}(2024)]%
        {jin2024llmindexer}
\bibfield{author}{\bibinfo{person}{Bowen Jin}, \bibinfo{person}{Hansi Zeng}, \bibinfo{person}{Guoyin Wang}, \bibinfo{person}{Xiusi Chen}, \bibinfo{person}{Tianxin Wei}, \bibinfo{person}{Ruirui Li}, \bibinfo{person}{Zhengyang Wang}, \bibinfo{person}{Zheng Li}, \bibinfo{person}{Yang Li}, \bibinfo{person}{Hanqing Lu}, \bibinfo{person}{Suhang Wang}, \bibinfo{person}{Jiawei Han}, {and} \bibinfo{person}{Xianfeng Tang}.} \bibinfo{year}{2024}\natexlab{}.
\newblock \showarticletitle{Language models as semantic indexers}. In \bibinfo{booktitle}{\emph{Proceedings of the 41st International Conference on Machine Learning}} (Vienna, Austria) \emph{(\bibinfo{series}{ICML'24})}. \bibinfo{publisher}{JMLR.org}, Article \bibinfo{articleno}{894}, \bibinfo{numpages}{16}~pages.
\newblock


\bibitem[Ju et~al\mbox{.}(2025)]%
        {ju2025grid}
\bibfield{author}{\bibinfo{person}{Clark~Mingxuan Ju}, \bibinfo{person}{Liam Collins}, \bibinfo{person}{Leonardo Neves}, \bibinfo{person}{Bhuvesh Kumar}, \bibinfo{person}{Louis~Yufeng Wang}, \bibinfo{person}{Tong Zhao}, {and} \bibinfo{person}{Neil Shah}.} \bibinfo{year}{2025}\natexlab{}.
\newblock \showarticletitle{Generative Recommendation with Semantic IDs: A Practitioner's Handbook}. In \bibinfo{booktitle}{\emph{Proceedings of the 34th ACM International Conference on Information and Knowledge Management (CIKM)}}.
\newblock


\bibitem[Kabbur et~al\mbox{.}(2013a)]%
        {Santosh2013FISM}
\bibfield{author}{\bibinfo{person}{Santosh Kabbur}, \bibinfo{person}{Xia Ning}, {and} \bibinfo{person}{George Karypis}.} \bibinfo{year}{2013}\natexlab{a}.
\newblock \showarticletitle{FISM: factored item similarity models for top-N recommender systems}. In \bibinfo{booktitle}{\emph{Proceedings of the 19th ACM SIGKDD International Conference on Knowledge Discovery and Data Mining}} (Chicago, Illinois, USA) \emph{(\bibinfo{series}{KDD '13})}. \bibinfo{publisher}{Association for Computing Machinery}, \bibinfo{address}{New York, NY, USA}, \bibinfo{pages}{659–667}.
\newblock
\showISBNx{9781450321747}
\href{https://doi.org/10.1145/2487575.2487589}{doi:\nolinkurl{10.1145/2487575.2487589}}


\bibitem[Kabbur et~al\mbox{.}(2013b)]%
        {Kabbur2013FISM}
\bibfield{author}{\bibinfo{person}{Santosh Kabbur}, \bibinfo{person}{Xia Ning}, {and} \bibinfo{person}{George Karypis}.} \bibinfo{year}{2013}\natexlab{b}.
\newblock \showarticletitle{FISM: factored item similarity models for top-N recommender systems}. In \bibinfo{booktitle}{\emph{Proceedings of the 19th ACM SIGKDD International Conference on Knowledge Discovery and Data Mining}} (Chicago, Illinois, USA) \emph{(\bibinfo{series}{KDD '13})}. \bibinfo{publisher}{Association for Computing Machinery}, \bibinfo{address}{New York, NY, USA}, \bibinfo{pages}{659–667}.
\newblock
\showISBNx{9781450321747}
\href{https://doi.org/10.1145/2487575.2487589}{doi:\nolinkurl{10.1145/2487575.2487589}}


\bibitem[Kang et~al\mbox{.}(2021)]%
        {kang2021learning}
\bibfield{author}{\bibinfo{person}{Wang-Cheng Kang}, \bibinfo{person}{Derek~Zhiyuan Cheng}, \bibinfo{person}{Tiansheng Yao}, \bibinfo{person}{Xinyang Yi}, \bibinfo{person}{Ting Chen}, \bibinfo{person}{Lichan Hong}, {and} \bibinfo{person}{Ed~H. Chi}.} \bibinfo{year}{2021}\natexlab{}.
\newblock \showarticletitle{Learning to Embed Categorical Features without Embedding Tables for Recommendation}. In \bibinfo{booktitle}{\emph{Proceedings of the 27th ACM SIGKDD Conference on Knowledge Discovery \& Data Mining}} (Virtual Event, Singapore) \emph{(\bibinfo{series}{KDD '21})}. \bibinfo{publisher}{Association for Computing Machinery}, \bibinfo{address}{New York, NY, USA}, \bibinfo{pages}{840–850}.
\newblock
\showISBNx{9781450383325}
\href{https://doi.org/10.1145/3447548.3467304}{doi:\nolinkurl{10.1145/3447548.3467304}}


\bibitem[Kang and McAuley(2018)]%
        {Kang2018SelfAttentiveSR}
\bibfield{author}{\bibinfo{person}{Wang-Cheng Kang} {and} \bibinfo{person}{Julian McAuley}.} \bibinfo{year}{2018}\natexlab{}.
\newblock \showarticletitle{Self-Attentive Sequential Recommendation}.
\newblock \bibinfo{journal}{\emph{2018 IEEE International Conference on Data Mining (ICDM)}} (\bibinfo{year}{2018}), \bibinfo{pages}{197--206}.
\newblock
\urldef\tempurl%
\url{https://api.semanticscholar.org/CorpusID:52127932}
\showURL{%
\tempurl}


\bibitem[Kaufman and Rousseeuw(1987)]%
        {Kaufman1987ClusteringBM}
\bibfield{author}{\bibinfo{person}{Leonard Kaufman} {and} \bibinfo{person}{Peter~J. Rousseeuw}.} \bibinfo{year}{1987}\natexlab{}.
\newblock \showarticletitle{Clustering by means of medoids}.
\newblock
\urldef\tempurl%
\url{https://api.semanticscholar.org/CorpusID:59662201}
\showURL{%
\tempurl}


\bibitem[Koh et~al\mbox{.}(2020)]%
        {Koh2020ConceptBM}
\bibfield{author}{\bibinfo{person}{Pang~Wei Koh}, \bibinfo{person}{Thao Nguyen}, \bibinfo{person}{Yew~Siang Tang}, \bibinfo{person}{Stephen Mussmann}, \bibinfo{person}{Emma Pierson}, \bibinfo{person}{Been Kim}, {and} \bibinfo{person}{Percy Liang}.} \bibinfo{year}{2020}\natexlab{}.
\newblock \showarticletitle{Concept Bottleneck Models}. In \bibinfo{booktitle}{\emph{Proceedings of the 37th International Conference on Machine Learning}} \emph{(\bibinfo{series}{Proceedings of Machine Learning Research}, Vol.~\bibinfo{volume}{119})}, \bibfield{editor}{\bibinfo{person}{Hal~Daumé III} {and} \bibinfo{person}{Aarti Singh}} (Eds.). \bibinfo{publisher}{PMLR}, \bibinfo{pages}{5338--5348}.
\newblock
\urldef\tempurl%
\url{https://proceedings.mlr.press/v119/koh20a.html}
\showURL{%
\tempurl}


\bibitem[Koren et~al\mbox{.}(2009)]%
        {Yehuda2009MF}
\bibfield{author}{\bibinfo{person}{Yehuda Koren}, \bibinfo{person}{Robert Bell}, {and} \bibinfo{person}{Chris Volinsky}.} \bibinfo{year}{2009}\natexlab{}.
\newblock \showarticletitle{Matrix Factorization Techniques for Recommender Systems}.
\newblock \bibinfo{journal}{\emph{Computer}} \bibinfo{volume}{42}, \bibinfo{number}{8} (\bibinfo{date}{Aug.} \bibinfo{year}{2009}), \bibinfo{pages}{30–37}.
\newblock
\showISSN{0018-9162}
\href{https://doi.org/10.1109/MC.2009.263}{doi:\nolinkurl{10.1109/MC.2009.263}}


\bibitem[Li et~al\mbox{.}(2018)]%
        {Raymond2018Towards}
\bibfield{author}{\bibinfo{person}{Raymond Li}, \bibinfo{person}{Samira Kahou}, \bibinfo{person}{Hannes Schulz}, \bibinfo{person}{Vincent Michalski}, \bibinfo{person}{Laurent Charlin}, {and} \bibinfo{person}{Chris Pal}.} \bibinfo{year}{2018}\natexlab{}.
\newblock \showarticletitle{Towards deep conversational recommendations}. In \bibinfo{booktitle}{\emph{Proceedings of the 32nd International Conference on Neural Information Processing Systems}} (Montr\'{e}al, Canada) \emph{(\bibinfo{series}{NIPS'18})}. \bibinfo{publisher}{Curran Associates Inc.}, \bibinfo{address}{Red Hook, NY, USA}, \bibinfo{pages}{9748–9758}.
\newblock


\bibitem[Liu et~al\mbox{.}(2025b)]%
        {liu2025improvingLLMPoweredRec}
\bibfield{author}{\bibinfo{person}{Jiahao Liu}, \bibinfo{person}{Xueshuo Yan}, \bibinfo{person}{Dongsheng Li}, \bibinfo{person}{Guangping Zhang}, \bibinfo{person}{Hansu Gu}, \bibinfo{person}{Peng Zhang}, \bibinfo{person}{Tun Lu}, \bibinfo{person}{Li Shang}, {and} \bibinfo{person}{Ning Gu}.} \bibinfo{year}{2025}\natexlab{b}.
\newblock \showarticletitle{Improving LLM-powered Recommendations with Personalized Information} \emph{(\bibinfo{series}{SIGIR '25})}. \bibinfo{publisher}{Association for Computing Machinery}, \bibinfo{address}{New York, NY, USA}, \bibinfo{pages}{2560–2565}.
\newblock
\showISBNx{9798400715921}
\href{https://doi.org/10.1145/3726302.3730211}{doi:\nolinkurl{10.1145/3726302.3730211}}


\bibitem[Liu et~al\mbox{.}(2024)]%
        {liu2024Once}
\bibfield{author}{\bibinfo{person}{Qijiong Liu}, \bibinfo{person}{Nuo Chen}, \bibinfo{person}{Tetsuya Sakai}, {and} \bibinfo{person}{Xiao-Ming Wu}.} \bibinfo{year}{2024}\natexlab{}.
\newblock \showarticletitle{ONCE: Boosting Content-based Recommendation with Both Open- and Closed-source Large Language Models}. In \bibinfo{booktitle}{\emph{Proceedings of the 17th ACM International Conference on Web Search and Data Mining}} (Merida, Mexico) \emph{(\bibinfo{series}{WSDM '24})}. \bibinfo{publisher}{Association for Computing Machinery}, \bibinfo{address}{New York, NY, USA}, \bibinfo{pages}{452–461}.
\newblock
\showISBNx{9798400703713}
\href{https://doi.org/10.1145/3616855.3635845}{doi:\nolinkurl{10.1145/3616855.3635845}}


\bibitem[Liu et~al\mbox{.}(2025a)]%
        {liu2025inferencecomputationscalingfeature}
\bibfield{author}{\bibinfo{person}{Weihao Liu}, \bibinfo{person}{Zhaocheng Du}, \bibinfo{person}{Haiyuan Zhao}, \bibinfo{person}{Wenbo Zhang}, \bibinfo{person}{Xiaoyan Zhao}, \bibinfo{person}{Gang Wang}, \bibinfo{person}{Zhenhua Dong}, {and} \bibinfo{person}{Jun Xu}.} \bibinfo{year}{2025}\natexlab{a}.
\newblock \bibinfo{title}{Inference Computation Scaling for Feature Augmentation in Recommendation Systems}.
\newblock
\showeprint[arxiv]{2502.16040}~[cs.IR]
\urldef\tempurl%
\url{https://arxiv.org/abs/2502.16040}
\showURL{%
\tempurl}


\bibitem[Madaan et~al\mbox{.}(2023)]%
        {madaan2023selfrefine}
\bibfield{author}{\bibinfo{person}{Aman Madaan}, \bibinfo{person}{Niket Tandon}, \bibinfo{person}{Prakhar Gupta}, \bibinfo{person}{Skyler Hallinan}, \bibinfo{person}{Luyu Gao}, \bibinfo{person}{Sarah Wiegreffe}, \bibinfo{person}{Uri Alon}, \bibinfo{person}{Nouha Dziri}, \bibinfo{person}{Shrimai Prabhumoye}, \bibinfo{person}{Yiming Yang}, \bibinfo{person}{Shashank Gupta}, \bibinfo{person}{Bodhisattwa~Prasad Majumder}, \bibinfo{person}{Katherine Hermann}, \bibinfo{person}{Sean Welleck}, \bibinfo{person}{Amir Yazdanbakhsh}, {and} \bibinfo{person}{Peter Clark}.} \bibinfo{year}{2023}\natexlab{}.
\newblock \showarticletitle{SELF-REFINE: iterative refinement with self-feedback}. In \bibinfo{booktitle}{\emph{Proceedings of the 37th International Conference on Neural Information Processing Systems}} (New Orleans, LA, USA) \emph{(\bibinfo{series}{NIPS '23})}. \bibinfo{publisher}{Curran Associates Inc.}, \bibinfo{address}{Red Hook, NY, USA}, Article \bibinfo{articleno}{2019}, \bibinfo{numpages}{61}~pages.
\newblock


\bibitem[Majumder et~al\mbox{.}(2024)]%
        {majumder2024clin}
\bibfield{author}{\bibinfo{person}{Bodhisattwa~Prasad Majumder}, \bibinfo{person}{Bhavana~Dalvi Mishra}, \bibinfo{person}{Peter Jansen}, \bibinfo{person}{Oyvind Tafjord}, \bibinfo{person}{Niket Tandon}, \bibinfo{person}{Li Zhang}, \bibinfo{person}{Chris Callison-Burch}, {and} \bibinfo{person}{Peter Clark}.} \bibinfo{year}{2024}\natexlab{}.
\newblock \showarticletitle{{CLIN}: A Continually Learning Language Agent for Rapid Task Adaptation and Generalization}. In \bibinfo{booktitle}{\emph{First Conference on Language Modeling}}.
\newblock
\urldef\tempurl%
\url{https://openreview.net/forum?id=xS6zx1aBI9}
\showURL{%
\tempurl}


\bibitem[McAuley et~al\mbox{.}(2015)]%
        {McAuley2015ImageBasedRec}
\bibfield{author}{\bibinfo{person}{Julian McAuley}, \bibinfo{person}{Christopher Targett}, \bibinfo{person}{Qinfeng Shi}, {and} \bibinfo{person}{Anton van~den Hengel}.} \bibinfo{year}{2015}\natexlab{}.
\newblock \showarticletitle{Image-Based Recommendations on Styles and Substitutes} \emph{(\bibinfo{series}{SIGIR '15})}. \bibinfo{publisher}{Association for Computing Machinery}, \bibinfo{address}{New York, NY, USA}, \bibinfo{pages}{43–52}.
\newblock
\showISBNx{9781450336215}
\href{https://doi.org/10.1145/2766462.2767755}{doi:\nolinkurl{10.1145/2766462.2767755}}


\bibitem[Mooney and Roy(2000)]%
        {mooney2000content}
\bibfield{author}{\bibinfo{person}{Raymond~J. Mooney} {and} \bibinfo{person}{Loriene Roy}.} \bibinfo{year}{2000}\natexlab{}.
\newblock \showarticletitle{Content-based book recommending using learning for text categorization}. In \bibinfo{booktitle}{\emph{Proceedings of the Fifth ACM Conference on Digital Libraries}} (San Antonio, Texas, USA) \emph{(\bibinfo{series}{DL '00})}. \bibinfo{publisher}{Association for Computing Machinery}, \bibinfo{address}{New York, NY, USA}, \bibinfo{pages}{195–204}.
\newblock
\showISBNx{158113231X}
\href{https://doi.org/10.1145/336597.336662}{doi:\nolinkurl{10.1145/336597.336662}}


\bibitem[Park(2025)]%
        {park2025LLMasUserSim}
\bibfield{author}{\bibinfo{person}{Choongwon Park}.} \bibinfo{year}{2025}\natexlab{}.
\newblock \showarticletitle{LLM as User Simulator: Towards Training News Recommender without Real User Interactions}. In \bibinfo{booktitle}{\emph{Proceedings of the 48th International ACM SIGIR Conference on Research and Development in Information Retrieval}} (Padua, Italy) \emph{(\bibinfo{series}{SIGIR '25})}. \bibinfo{publisher}{Association for Computing Machinery}, \bibinfo{address}{New York, NY, USA}, \bibinfo{pages}{3080–3084}.
\newblock
\showISBNx{9798400715921}
\href{https://doi.org/10.1145/3726302.3730224}{doi:\nolinkurl{10.1145/3726302.3730224}}


\bibitem[Pham et~al\mbox{.}(2024)]%
        {pham-etal-2024-topicgpt}
\bibfield{author}{\bibinfo{person}{Chau~Minh Pham}, \bibinfo{person}{Alexander Hoyle}, \bibinfo{person}{Simeng Sun}, \bibinfo{person}{Philip Resnik}, {and} \bibinfo{person}{Mohit Iyyer}.} \bibinfo{year}{2024}\natexlab{}.
\newblock \showarticletitle{{T}opic{GPT}: A Prompt-based Topic Modeling Framework}. In \bibinfo{booktitle}{\emph{Proceedings of the 2024 Conference of the North American Chapter of the Association for Computational Linguistics: Human Language Technologies (Volume 1: Long Papers)}}, \bibfield{editor}{\bibinfo{person}{Kevin Duh}, \bibinfo{person}{Helena Gomez}, {and} \bibinfo{person}{Steven Bethard}} (Eds.). \bibinfo{publisher}{Association for Computational Linguistics}, \bibinfo{address}{Mexico City, Mexico}, \bibinfo{pages}{2956--2984}.
\newblock
\href{https://doi.org/10.18653/v1/2024.naacl-long.164}{doi:\nolinkurl{10.18653/v1/2024.naacl-long.164}}


\bibitem[Pope et~al\mbox{.}(2023)]%
        {pope2023efficient}
\bibfield{author}{\bibinfo{person}{Reiner Pope}, \bibinfo{person}{Sholto Douglas}, \bibinfo{person}{Aakanksha Chowdhery}, \bibinfo{person}{Jacob Devlin}, \bibinfo{person}{James Bradbury}, \bibinfo{person}{Jonathan Heek}, \bibinfo{person}{Kefan Xiao}, \bibinfo{person}{Shivani Agrawal}, {and} \bibinfo{person}{Jeff Dean}.} \bibinfo{year}{2023}\natexlab{}.
\newblock \showarticletitle{Efficiently Scaling Transformer Inference}. In \bibinfo{booktitle}{\emph{Proceedings of Machine Learning and Systems}}, \bibfield{editor}{\bibinfo{person}{D.~Song}, \bibinfo{person}{M.~Carbin}, {and} \bibinfo{person}{T.~Chen}} (Eds.), Vol.~\bibinfo{volume}{5}. \bibinfo{publisher}{Curan}, \bibinfo{pages}{606--624}.
\newblock


\bibitem[Rajput et~al\mbox{.}(2023)]%
        {rajput2023recommender}
\bibfield{author}{\bibinfo{person}{Shashank Rajput}, \bibinfo{person}{Nikhil Mehta}, \bibinfo{person}{Anima Singh}, \bibinfo{person}{Raghunandan~Hulikal Keshavan}, \bibinfo{person}{Trung Vu}, \bibinfo{person}{Lukasz Heldt}, \bibinfo{person}{Lichan Hong}, \bibinfo{person}{Yi Tay}, \bibinfo{person}{Vinh~Q. Tran}, \bibinfo{person}{Jonah Samost}, \bibinfo{person}{Maciej Kula}, \bibinfo{person}{Ed~H. Chi}, {and} \bibinfo{person}{Maheswaran Sathiamoorthy}.} \bibinfo{year}{2023}\natexlab{}.
\newblock \showarticletitle{Recommender Systems with Generative Retrieval}. In \bibinfo{booktitle}{\emph{Thirty-seventh Conference on Neural Information Processing Systems}}.
\newblock
\urldef\tempurl%
\url{https://openreview.net/forum?id=BJ0fQUU32w}
\showURL{%
\tempurl}


\bibitem[Ren et~al\mbox{.}(2024)]%
        {Ren2024EnhancingSeq}
\bibfield{author}{\bibinfo{person}{Yankun Ren}, \bibinfo{person}{Zhongde Chen}, \bibinfo{person}{Xinxing Yang}, \bibinfo{person}{Longfei Li}, \bibinfo{person}{Cong Jiang}, \bibinfo{person}{Lei Cheng}, \bibinfo{person}{Bo Zhang}, \bibinfo{person}{Linjian Mo}, {and} \bibinfo{person}{Jun Zhou}.} \bibinfo{year}{2024}\natexlab{}.
\newblock \showarticletitle{Enhancing Sequential Recommenders with Augmented Knowledge from Aligned Large Language Models}. In \bibinfo{booktitle}{\emph{Proceedings of the 47th International ACM SIGIR Conference on Research and Development in Information Retrieval}} (Washington DC, USA) \emph{(\bibinfo{series}{SIGIR '24})}. \bibinfo{publisher}{Association for Computing Machinery}, \bibinfo{address}{New York, NY, USA}, \bibinfo{pages}{345–354}.
\newblock
\showISBNx{9798400704314}
\href{https://doi.org/10.1145/3626772.3657782}{doi:\nolinkurl{10.1145/3626772.3657782}}


\bibitem[Rendle et~al\mbox{.}(2009)]%
        {Steffen2009BPR}
\bibfield{author}{\bibinfo{person}{Steffen Rendle}, \bibinfo{person}{Christoph Freudenthaler}, \bibinfo{person}{Zeno Gantner}, {and} \bibinfo{person}{Lars Schmidt-Thieme}.} \bibinfo{year}{2009}\natexlab{}.
\newblock \showarticletitle{BPR: Bayesian personalized ranking from implicit feedback}. In \bibinfo{booktitle}{\emph{Proceedings of the Twenty-Fifth Conference on Uncertainty in Artificial Intelligence}} (Montreal, Quebec, Canada) \emph{(\bibinfo{series}{UAI '09})}. \bibinfo{publisher}{AUAI Press}, \bibinfo{address}{Arlington, Virginia, USA}, \bibinfo{pages}{452–461}.
\newblock
\showISBNx{9780974903958}


\bibitem[Sharma et~al\mbox{.}(2015)]%
        {sharma2015featurebased}
\bibfield{author}{\bibinfo{person}{Mohit Sharma}, \bibinfo{person}{Jiayu Zhou}, \bibinfo{person}{Junling Hu}, {and} \bibinfo{person}{George Karypis}.} \bibinfo{year}{2015}\natexlab{}.
\newblock \showarticletitle{Feature-based factorized Bilinear Similarity Model for Cold-Start Top-n Item Recommendation}. In \bibinfo{booktitle}{\emph{Proceedings of the 2015 SIAM International Conference on Data Mining}}. \bibinfo{publisher}{Society for Industrial and Applied Mathematics}, \bibinfo{pages}{190–198}.
\newblock
\href{https://doi.org/10.1137/1.9781611974010.22}{doi:\nolinkurl{10.1137/1.9781611974010.22}}


\bibitem[Shi et~al\mbox{.}(2025)]%
        {shi2025mattersllmbasedfeatureextractor}
\bibfield{author}{\bibinfo{person}{Kainan Shi}, \bibinfo{person}{Peilin Zhou}, \bibinfo{person}{Ge Wang}, \bibinfo{person}{Han Ding}, {and} \bibinfo{person}{Fei Wang}.} \bibinfo{year}{2025}\natexlab{}.
\newblock \bibinfo{title}{What Matters in LLM-Based Feature Extractor for Recommender? A Systematic Analysis of Prompts, Models, and Adaptation}.
\newblock
\showeprint[arxiv]{2509.14979}~[cs.IR]
\urldef\tempurl%
\url{https://arxiv.org/abs/2509.14979}
\showURL{%
\tempurl}


\bibitem[Shinn et~al\mbox{.}(2023)]%
        {shinn2023reflection}
\bibfield{author}{\bibinfo{person}{Noah Shinn}, \bibinfo{person}{Federico Cassano}, \bibinfo{person}{Ashwin Gopinath}, \bibinfo{person}{Karthik Narasimhan}, {and} \bibinfo{person}{Shunyu Yao}.} \bibinfo{year}{2023}\natexlab{}.
\newblock \showarticletitle{Reflexion: language agents with verbal reinforcement learning}. In \bibinfo{booktitle}{\emph{Proceedings of the 37th International Conference on Neural Information Processing Systems}} (New Orleans, LA, USA) \emph{(\bibinfo{series}{NIPS '23})}. \bibinfo{publisher}{Curran Associates Inc.}, \bibinfo{address}{Red Hook, NY, USA}, Article \bibinfo{articleno}{377}, \bibinfo{numpages}{19}~pages.
\newblock


\bibitem[Shu et~al\mbox{.}(2023)]%
        {shu2023rahrecsysassistanthumanhumancenteredrecommendation}
\bibfield{author}{\bibinfo{person}{Yubo Shu}, \bibinfo{person}{Haonan Zhang}, \bibinfo{person}{Hansu Gu}, \bibinfo{person}{Peng Zhang}, \bibinfo{person}{Tun Lu}, \bibinfo{person}{Dongsheng Li}, {and} \bibinfo{person}{Ning Gu}.} \bibinfo{year}{2023}\natexlab{}.
\newblock \bibinfo{title}{RAH! RecSys-Assistant-Human: A Human-Centered Recommendation Framework with LLM Agents}.
\newblock
\showeprint[arxiv]{2308.09904}~[cs.IR]
\urldef\tempurl%
\url{https://arxiv.org/abs/2308.09904}
\showURL{%
\tempurl}


\bibitem[Singh et~al\mbox{.}(2024)]%
        {Anima2024Better}
\bibfield{author}{\bibinfo{person}{Anima Singh}, \bibinfo{person}{Trung Vu}, \bibinfo{person}{Nikhil Mehta}, \bibinfo{person}{Raghunandan Keshavan}, \bibinfo{person}{Maheswaran Sathiamoorthy}, \bibinfo{person}{Yilin Zheng}, \bibinfo{person}{Lichan Hong}, \bibinfo{person}{Lukasz Heldt}, \bibinfo{person}{Li Wei}, \bibinfo{person}{Devansh Tandon}, \bibinfo{person}{Ed Chi}, {and} \bibinfo{person}{Xinyang Yi}.} \bibinfo{year}{2024}\natexlab{}.
\newblock \showarticletitle{Better Generalization with Semantic IDs: A Case Study in Ranking for Recommendations}. In \bibinfo{booktitle}{\emph{Proceedings of the 18th ACM Conference on Recommender Systems}} (Bari, Italy) \emph{(\bibinfo{series}{RecSys '24})}. \bibinfo{publisher}{Association for Computing Machinery}, \bibinfo{address}{New York, NY, USA}, \bibinfo{pages}{1039–1044}.
\newblock
\showISBNx{9798400705052}
\href{https://doi.org/10.1145/3640457.3688190}{doi:\nolinkurl{10.1145/3640457.3688190}}


\bibitem[Song et~al\mbox{.}(2019)]%
        {song2019AutoInt}
\bibfield{author}{\bibinfo{person}{Weiping Song}, \bibinfo{person}{Chence Shi}, \bibinfo{person}{Zhiping Xiao}, \bibinfo{person}{Zhijian Duan}, \bibinfo{person}{Yewen Xu}, \bibinfo{person}{Ming Zhang}, {and} \bibinfo{person}{Jian Tang}.} \bibinfo{year}{2019}\natexlab{}.
\newblock \showarticletitle{AutoInt: Automatic Feature Interaction Learning via Self-Attentive Neural Networks} \emph{(\bibinfo{series}{CIKM '19})}. \bibinfo{publisher}{Association for Computing Machinery}, \bibinfo{address}{New York, NY, USA}, \bibinfo{pages}{1161–1170}.
\newblock
\showISBNx{9781450369763}
\href{https://doi.org/10.1145/3357384.3357925}{doi:\nolinkurl{10.1145/3357384.3357925}}


\bibitem[Sun et~al\mbox{.}(2025)]%
        {sun2025concept}
\bibfield{author}{\bibinfo{person}{Chung-En Sun}, \bibinfo{person}{Tuomas Oikarinen}, \bibinfo{person}{Berk Ustun}, {and} \bibinfo{person}{Tsui-Wei Weng}.} \bibinfo{year}{2025}\natexlab{}.
\newblock \showarticletitle{Concept Bottleneck Large Language Models}. In \bibinfo{booktitle}{\emph{The Thirteenth International Conference on Learning Representations}}.
\newblock
\urldef\tempurl%
\url{https://openreview.net/forum?id=RC5FPYVQaH}
\showURL{%
\tempurl}


\bibitem[Sun et~al\mbox{.}(2019)]%
        {fei2019bert4rec}
\bibfield{author}{\bibinfo{person}{Fei Sun}, \bibinfo{person}{Jun Liu}, \bibinfo{person}{Jian Wu}, \bibinfo{person}{Changhua Pei}, \bibinfo{person}{Xiao Lin}, \bibinfo{person}{Wenwu Ou}, {and} \bibinfo{person}{Peng Jiang}.} \bibinfo{year}{2019}\natexlab{}.
\newblock \showarticletitle{BERT4Rec: Sequential Recommendation with Bidirectional Encoder Representations from Transformer}. In \bibinfo{booktitle}{\emph{Proceedings of the 28th ACM International Conference on Information and Knowledge Management}} (Beijing, China) \emph{(\bibinfo{series}{CIKM '19})}. \bibinfo{publisher}{Association for Computing Machinery}, \bibinfo{address}{New York, NY, USA}, \bibinfo{pages}{1441–1450}.
\newblock
\showISBNx{9781450369763}
\href{https://doi.org/10.1145/3357384.3357895}{doi:\nolinkurl{10.1145/3357384.3357895}}


\bibitem[Thakkar and Yadav(2024)]%
        {thakkar2024personalizedrecommendationsystemsusing}
\bibfield{author}{\bibinfo{person}{Param Thakkar} {and} \bibinfo{person}{Anushka Yadav}.} \bibinfo{year}{2024}\natexlab{}.
\newblock \bibinfo{title}{Personalized Recommendation Systems using Multimodal, Autonomous, Multi Agent Systems}.
\newblock
\showeprint[arxiv]{2410.19855}~[cs.IR]
\urldef\tempurl%
\url{https://arxiv.org/abs/2410.19855}
\showURL{%
\tempurl}


\bibitem[Valizadeh et~al\mbox{.}(2025)]%
        {valizadeh2025languagemodelssemanticaugmenters}
\bibfield{author}{\bibinfo{person}{Mahsa Valizadeh}, \bibinfo{person}{Xiangjue Dong}, \bibinfo{person}{Rui Tuo}, {and} \bibinfo{person}{James Caverlee}.} \bibinfo{year}{2025}\natexlab{}.
\newblock \bibinfo{title}{Language Models as Semantic Augmenters for Sequential Recommenders}.
\newblock
\showeprint[arxiv]{2510.18046}~[cs.CL]
\urldef\tempurl%
\url{https://arxiv.org/abs/2510.18046}
\showURL{%
\tempurl}


\bibitem[van~den Oord et~al\mbox{.}(2017)]%
        {van2017neuraldiscrete}
\bibfield{author}{\bibinfo{person}{Aaron van~den Oord}, \bibinfo{person}{Oriol Vinyals}, {and} \bibinfo{person}{Koray Kavukcuoglu}.} \bibinfo{year}{2017}\natexlab{}.
\newblock \showarticletitle{Neural discrete representation learning}. In \bibinfo{booktitle}{\emph{Proceedings of the 31st International Conference on Neural Information Processing Systems}} (Long Beach, California, USA) \emph{(\bibinfo{series}{NIPS'17})}. \bibinfo{publisher}{Curran Associates Inc.}, \bibinfo{address}{Red Hook, NY, USA}, \bibinfo{pages}{6309–6318}.
\newblock
\showISBNx{9781510860964}


\bibitem[Wan et~al\mbox{.}(2024)]%
        {Wan2024TnTLLM}
\bibfield{author}{\bibinfo{person}{Mengting Wan}, \bibinfo{person}{Tara Safavi}, \bibinfo{person}{Sujay~Kumar Jauhar}, \bibinfo{person}{Yujin Kim}, \bibinfo{person}{Scott Counts}, \bibinfo{person}{Jennifer Neville}, \bibinfo{person}{Siddharth Suri}, \bibinfo{person}{Chirag Shah}, \bibinfo{person}{Ryen~W. White}, \bibinfo{person}{Longqi Yang}, \bibinfo{person}{Reid Andersen}, \bibinfo{person}{Georg Buscher}, \bibinfo{person}{Dhruv Joshi}, {and} \bibinfo{person}{Nagu Rangan}.} \bibinfo{year}{2024}\natexlab{}.
\newblock \showarticletitle{TnT-LLM: Text Mining at Scale with Large Language Models}. In \bibinfo{booktitle}{\emph{Proceedings of the 30th ACM SIGKDD Conference on Knowledge Discovery and Data Mining}} (Barcelona, Spain) \emph{(\bibinfo{series}{KDD '24})}. \bibinfo{publisher}{Association for Computing Machinery}, \bibinfo{address}{New York, NY, USA}, \bibinfo{pages}{5836–5847}.
\newblock
\showISBNx{9798400704901}
\href{https://doi.org/10.1145/3637528.3671647}{doi:\nolinkurl{10.1145/3637528.3671647}}


\bibitem[Wang et~al\mbox{.}(2024c)]%
        {wang2024llmasda}
\bibfield{author}{\bibinfo{person}{Jianling Wang}, \bibinfo{person}{Haokai Lu}, \bibinfo{person}{James Caverlee}, \bibinfo{person}{Ed~H. Chi}, {and} \bibinfo{person}{Minmin Chen}.} \bibinfo{year}{2024}\natexlab{c}.
\newblock \showarticletitle{Large Language Models as Data Augmenters for Cold-Start Item Recommendation}. In \bibinfo{booktitle}{\emph{Companion Proceedings of the ACM Web Conference 2024}} (Singapore, Singapore) \emph{(\bibinfo{series}{WWW '24})}. \bibinfo{publisher}{Association for Computing Machinery}, \bibinfo{address}{New York, NY, USA}, \bibinfo{pages}{726–729}.
\newblock
\showISBNx{9798400701726}
\href{https://doi.org/10.1145/3589335.3651532}{doi:\nolinkurl{10.1145/3589335.3651532}}


\bibitem[Wang et~al\mbox{.}(2024d)]%
        {wang2024llmDA}
\bibfield{author}{\bibinfo{person}{Jianling Wang}, \bibinfo{person}{Haokai Lu}, \bibinfo{person}{James Caverlee}, \bibinfo{person}{Ed~H. Chi}, {and} \bibinfo{person}{Minmin Chen}.} \bibinfo{year}{2024}\natexlab{d}.
\newblock \showarticletitle{Large Language Models as Data Augmenters for Cold-Start Item Recommendation}. In \bibinfo{booktitle}{\emph{Companion Proceedings of the ACM Web Conference 2024}} (Singapore, Singapore) \emph{(\bibinfo{series}{WWW '24})}. \bibinfo{publisher}{Association for Computing Machinery}, \bibinfo{address}{New York, NY, USA}, \bibinfo{pages}{726–729}.
\newblock
\showISBNx{9798400701726}
\href{https://doi.org/10.1145/3589335.3651532}{doi:\nolinkurl{10.1145/3589335.3651532}}


\bibitem[Wang et~al\mbox{.}(2024f)]%
        {wang2024userbehavsim}
\bibfield{author}{\bibinfo{person}{Lei Wang}, \bibinfo{person}{Jingsen Zhang}, \bibinfo{person}{Hao Yang}, \bibinfo{person}{Zhiyuan Chen}, \bibinfo{person}{Jiakai Tang}, \bibinfo{person}{Zeyu Zhang}, \bibinfo{person}{Xu Chen}, \bibinfo{person}{Yankai Lin}, \bibinfo{person}{Ruihua Song}, \bibinfo{person}{Wayne~Xin Zhao}, \bibinfo{person}{Jun Xu}, \bibinfo{person}{Zhicheng Dou}, \bibinfo{person}{Jun Wang}, {and} \bibinfo{person}{Ji-Rong Wen}.} \bibinfo{year}{2024}\natexlab{f}.
\newblock \bibinfo{title}{User Behavior Simulation with Large Language Model-based Agents}.
\newblock
\showeprint[arxiv]{2306.02552}~[cs.IR]


\bibitem[Wang et~al\mbox{.}(2017)]%
        {wang2017DCN}
\bibfield{author}{\bibinfo{person}{Ruoxi Wang}, \bibinfo{person}{Bin Fu}, \bibinfo{person}{Gang Fu}, {and} \bibinfo{person}{Mingliang Wang}.} \bibinfo{year}{2017}\natexlab{}.
\newblock \showarticletitle{Deep \& Cross Network for Ad Click Predictions}. In \bibinfo{booktitle}{\emph{Proceedings of the ADKDD'17}} (Halifax, NS, Canada) \emph{(\bibinfo{series}{ADKDD'17})}. \bibinfo{publisher}{Association for Computing Machinery}, \bibinfo{address}{New York, NY, USA}, Article \bibinfo{articleno}{12}, \bibinfo{numpages}{7}~pages.
\newblock
\showISBNx{9781450351942}
\href{https://doi.org/10.1145/3124749.3124754}{doi:\nolinkurl{10.1145/3124749.3124754}}


\bibitem[Wang et~al\mbox{.}(2021)]%
        {Wang2021DCNV2}
\bibfield{author}{\bibinfo{person}{Ruoxi Wang}, \bibinfo{person}{Rakesh Shivanna}, \bibinfo{person}{Derek Cheng}, \bibinfo{person}{Sagar Jain}, \bibinfo{person}{Dong Lin}, \bibinfo{person}{Lichan Hong}, {and} \bibinfo{person}{Ed Chi}.} \bibinfo{year}{2021}\natexlab{}.
\newblock \showarticletitle{DCN V2: Improved Deep \& Cross Network and Practical Lessons for Web-scale Learning to Rank Systems}. In \bibinfo{booktitle}{\emph{Proceedings of the Web Conference 2021}} (Ljubljana, Slovenia) \emph{(\bibinfo{series}{WWW '21})}. \bibinfo{publisher}{Association for Computing Machinery}, \bibinfo{address}{New York, NY, USA}, \bibinfo{pages}{1785–1797}.
\newblock
\showISBNx{9781450383127}
\href{https://doi.org/10.1145/3442381.3450078}{doi:\nolinkurl{10.1145/3442381.3450078}}


\bibitem[Wang et~al\mbox{.}(2024a)]%
        {wang2024learnableitemtokenizationgenerative}
\bibfield{author}{\bibinfo{person}{Wenjie Wang}, \bibinfo{person}{Honghui Bao}, \bibinfo{person}{Xinyu Lin}, \bibinfo{person}{Jizhi Zhang}, \bibinfo{person}{Yongqi Li}, \bibinfo{person}{Fuli Feng}, \bibinfo{person}{See-Kiong Ng}, {and} \bibinfo{person}{Tat-Seng Chua}.} \bibinfo{year}{2024}\natexlab{a}.
\newblock \showarticletitle{Learnable Item Tokenization for Generative Recommendation}. In \bibinfo{booktitle}{\emph{International Conference on Information and Knowledge Management}}.
\newblock


\bibitem[Wang et~al\mbox{.}(2023)]%
        {wang-etal-2023-rethinking-evaluation}
\bibfield{author}{\bibinfo{person}{Xiaolei Wang}, \bibinfo{person}{Xinyu Tang}, \bibinfo{person}{Xin Zhao}, \bibinfo{person}{Jingyuan Wang}, {and} \bibinfo{person}{Ji-Rong Wen}.} \bibinfo{year}{2023}\natexlab{}.
\newblock \showarticletitle{Rethinking the Evaluation for Conversational Recommendation in the Era of Large Language Models}. In \bibinfo{booktitle}{\emph{Proceedings of the 2023 Conference on Empirical Methods in Natural Language Processing}}, \bibfield{editor}{\bibinfo{person}{Houda Bouamor}, \bibinfo{person}{Juan Pino}, {and} \bibinfo{person}{Kalika Bali}} (Eds.). \bibinfo{publisher}{Association for Computational Linguistics}, \bibinfo{address}{Singapore}, \bibinfo{pages}{10052--10065}.
\newblock
\href{https://doi.org/10.18653/v1/2023.emnlp-main.621}{doi:\nolinkurl{10.18653/v1/2023.emnlp-main.621}}


\bibitem[Wang et~al\mbox{.}(2022)]%
        {wang2022towards}
\bibfield{author}{\bibinfo{person}{Xiaolei Wang}, \bibinfo{person}{Kun Zhou}, \bibinfo{person}{Ji-Rong Wen}, {and} \bibinfo{person}{Wayne~Xin Zhao}.} \bibinfo{year}{2022}\natexlab{}.
\newblock \showarticletitle{Towards Unified Conversational Recommender Systems via Knowledge-Enhanced Prompt Learning}. In \bibinfo{booktitle}{\emph{Proceedings of the 28th ACM SIGKDD Conference on Knowledge Discovery and Data Mining}}. \bibinfo{pages}{1929--1937}.
\newblock


\bibitem[Wang et~al\mbox{.}(2024b)]%
        {wang2024recmind}
\bibfield{author}{\bibinfo{person}{Yancheng Wang}, \bibinfo{person}{Ziyan Jiang}, \bibinfo{person}{Zheng Chen}, \bibinfo{person}{Fan Yang}, \bibinfo{person}{Yingxue Zhou}, \bibinfo{person}{Eunah Cho}, \bibinfo{person}{Xing Fan}, \bibinfo{person}{Yanbin Lu}, \bibinfo{person}{Xiaojiang Huang}, {and} \bibinfo{person}{Yingzhen Yang}.} \bibinfo{year}{2024}\natexlab{b}.
\newblock \showarticletitle{{R}ec{M}ind: Large Language Model Powered Agent For Recommendation}. In \bibinfo{booktitle}{\emph{Findings of the Association for Computational Linguistics: NAACL 2024}}, \bibfield{editor}{\bibinfo{person}{Kevin Duh}, \bibinfo{person}{Helena Gomez}, {and} \bibinfo{person}{Steven Bethard}} (Eds.). \bibinfo{publisher}{Association for Computational Linguistics}, \bibinfo{address}{Mexico City, Mexico}, \bibinfo{pages}{4351--4364}.
\newblock
\href{https://doi.org/10.18653/v1/2024.findings-naacl.271}{doi:\nolinkurl{10.18653/v1/2024.findings-naacl.271}}


\bibitem[Wang et~al\mbox{.}(2024e)]%
        {wang2024colla}
\bibfield{author}{\bibinfo{person}{Yidan Wang}, \bibinfo{person}{Zhaochun Ren}, \bibinfo{person}{Weiwei Sun}, \bibinfo{person}{Jiyuan Yang}, \bibinfo{person}{Zhixiang Liang}, \bibinfo{person}{Xin Chen}, \bibinfo{person}{Ruobing Xie}, \bibinfo{person}{Su Yan}, \bibinfo{person}{Xu Zhang}, \bibinfo{person}{Pengjie Ren}, \bibinfo{person}{Zhumin Chen}, {and} \bibinfo{person}{Xin Xin}.} \bibinfo{year}{2024}\natexlab{e}.
\newblock \showarticletitle{Content-Based Collaborative Generation for Recommender Systems}. In \bibinfo{booktitle}{\emph{Proceedings of the 33rd ACM International Conference on Information and Knowledge Management}} (Boise, ID, USA) \emph{(\bibinfo{series}{CIKM '24})}. \bibinfo{publisher}{Association for Computing Machinery}, \bibinfo{address}{New York, NY, USA}, \bibinfo{pages}{2420–2430}.
\newblock
\showISBNx{9798400704369}
\href{https://doi.org/10.1145/3627673.3679692}{doi:\nolinkurl{10.1145/3627673.3679692}}


\bibitem[Wartena et~al\mbox{.}(2010)]%
        {christian2010selecting}
\bibfield{author}{\bibinfo{person}{Christian Wartena}, \bibinfo{person}{Wout Slakhorst}, {and} \bibinfo{person}{Martin Wibbels}.} \bibinfo{year}{2010}\natexlab{}.
\newblock \showarticletitle{Selecting keywords for content based recommendation}. In \bibinfo{booktitle}{\emph{Proceedings of the 19th ACM International Conference on Information and Knowledge Management}} (Toronto, ON, Canada) \emph{(\bibinfo{series}{CIKM '10})}. \bibinfo{publisher}{Association for Computing Machinery}, \bibinfo{address}{New York, NY, USA}, \bibinfo{pages}{1533–1536}.
\newblock
\showISBNx{9781450300995}
\href{https://doi.org/10.1145/1871437.1871665}{doi:\nolinkurl{10.1145/1871437.1871665}}


\bibitem[Xia et~al\mbox{.}(2026)]%
        {xia2026multiagentcollaborativefilteringorchestrating}
\bibfield{author}{\bibinfo{person}{Yu Xia}, \bibinfo{person}{Sungchul Kim}, \bibinfo{person}{Tong Yu}, \bibinfo{person}{Ryan~A. Rossi}, {and} \bibinfo{person}{Julian McAuley}.} \bibinfo{year}{2026}\natexlab{}.
\newblock \bibinfo{title}{Multi-Agent Collaborative Filtering: Orchestrating Users and Items for Agentic Recommendations}.
\newblock
\showeprint[arxiv]{2511.18413}~[cs.CL]
\urldef\tempurl%
\url{https://arxiv.org/abs/2511.18413}
\showURL{%
\tempurl}


\bibitem[Xie et~al\mbox{.}(2025)]%
        {xie-etal-2025-latent}
\bibfield{author}{\bibinfo{person}{Zhouhang Xie}, \bibinfo{person}{Tushar Khot}, \bibinfo{person}{Bhavana Dalvi~Mishra}, \bibinfo{person}{Harshit Surana}, \bibinfo{person}{Julian McAuley}, \bibinfo{person}{Peter Clark}, {and} \bibinfo{person}{Bodhisattwa~Prasad Majumder}.} \bibinfo{year}{2025}\natexlab{}.
\newblock \showarticletitle{Latent Factor Models Meets Instructions: Goal-conditioned Latent Factor Discovery without Task Supervision}. In \bibinfo{booktitle}{\emph{Proceedings of the 2025 Conference of the Nations of the Americas Chapter of the Association for Computational Linguistics: Human Language Technologies (Volume 1: Long Papers)}}, \bibfield{editor}{\bibinfo{person}{Luis Chiruzzo}, \bibinfo{person}{Alan Ritter}, {and} \bibinfo{person}{Lu~Wang}} (Eds.). \bibinfo{publisher}{Association for Computational Linguistics}, \bibinfo{address}{Albuquerque, New Mexico}, \bibinfo{pages}{11114--11134}.
\newblock
\showISBNx{979-8-89176-189-6}
\href{https://doi.org/10.18653/v1/2025.naacl-long.554}{doi:\nolinkurl{10.18653/v1/2025.naacl-long.554}}


\bibitem[Xie et~al\mbox{.}(2024)]%
        {xie-etal-2024-shot-dialogue}
\bibfield{author}{\bibinfo{person}{Zhouhang Xie}, \bibinfo{person}{Bodhisattwa~Prasad Majumder}, \bibinfo{person}{Mengjie Zhao}, \bibinfo{person}{Yoshinori Maeda}, \bibinfo{person}{Keiichi Yamada}, \bibinfo{person}{Hiromi Wakaki}, {and} \bibinfo{person}{Julian McAuley}.} \bibinfo{year}{2024}\natexlab{}.
\newblock \showarticletitle{Few-shot Dialogue Strategy Learning for Motivational Interviewing via Inductive Reasoning}. In \bibinfo{booktitle}{\emph{Findings of the Association for Computational Linguistics: ACL 2024}}, \bibfield{editor}{\bibinfo{person}{Lun-Wei Ku}, \bibinfo{person}{Andre Martins}, {and} \bibinfo{person}{Vivek Srikumar}} (Eds.). \bibinfo{publisher}{Association for Computational Linguistics}, \bibinfo{address}{Bangkok, Thailand}, \bibinfo{pages}{13207--13219}.
\newblock
\href{https://doi.org/10.18653/v1/2024.findings-acl.782}{doi:\nolinkurl{10.18653/v1/2024.findings-acl.782}}


\bibitem[Yada and Yamana(2024)]%
        {yada2024newsrecommendationcategorydescription}
\bibfield{author}{\bibinfo{person}{Yuki Yada} {and} \bibinfo{person}{Hayato Yamana}.} \bibinfo{year}{2024}\natexlab{}.
\newblock \bibinfo{title}{News Recommendation with Category Description by a Large Language Model}.
\newblock
\showeprint[arxiv]{2405.13007}~[cs.CL]
\urldef\tempurl%
\url{https://arxiv.org/abs/2405.13007}
\showURL{%
\tempurl}


\bibitem[Zhai et~al\mbox{.}(2024)]%
        {zhao2024hstu}
\bibfield{author}{\bibinfo{person}{Jiaqi Zhai}, \bibinfo{person}{Lucy Liao}, \bibinfo{person}{Xing Liu}, \bibinfo{person}{Yueming Wang}, \bibinfo{person}{Rui Li}, \bibinfo{person}{Xuan Cao}, \bibinfo{person}{Leon Gao}, \bibinfo{person}{Zhaojie Gong}, \bibinfo{person}{Fangda Gu}, \bibinfo{person}{Jiayuan He}, \bibinfo{person}{Yinghai Lu}, {and} \bibinfo{person}{Yu Shi}.} \bibinfo{year}{2024}\natexlab{}.
\newblock \showarticletitle{Actions Speak Louder than Words: Trillion-Parameter Sequential Transducers for Generative Recommendations}. In \bibinfo{booktitle}{\emph{Proceedings of the 41st International Conference on Machine Learning}} \emph{(\bibinfo{series}{Proceedings of Machine Learning Research}, Vol.~\bibinfo{volume}{235})}, \bibfield{editor}{\bibinfo{person}{Ruslan Salakhutdinov}, \bibinfo{person}{Zico Kolter}, \bibinfo{person}{Katherine Heller}, \bibinfo{person}{Adrian Weller}, \bibinfo{person}{Nuria Oliver}, \bibinfo{person}{Jonathan Scarlett}, {and} \bibinfo{person}{Felix Berkenkamp}} (Eds.). \bibinfo{publisher}{PMLR}, \bibinfo{pages}{58484--58509}.
\newblock
\urldef\tempurl%
\url{https://proceedings.mlr.press/v235/zhai24a.html}
\showURL{%
\tempurl}


\bibitem[Zhang et~al\mbox{.}(2024b)]%
        {Zhang2024NoteLLM}
\bibfield{author}{\bibinfo{person}{Chao Zhang}, \bibinfo{person}{Shiwei Wu}, \bibinfo{person}{Haoxin Zhang}, \bibinfo{person}{Tong Xu}, \bibinfo{person}{Yan Gao}, \bibinfo{person}{Yao Hu}, {and} \bibinfo{person}{Enhong Chen}.} \bibinfo{year}{2024}\natexlab{b}.
\newblock \showarticletitle{NoteLLM: A Retrievable Large Language Model for Note Recommendation} \emph{(\bibinfo{series}{WWW '24})}. \bibinfo{publisher}{Association for Computing Machinery}, \bibinfo{address}{New York, NY, USA}, \bibinfo{pages}{170–179}.
\newblock
\showISBNx{9798400701726}
\href{https://doi.org/10.1145/3589335.3648314}{doi:\nolinkurl{10.1145/3589335.3648314}}


\bibitem[Zhang et~al\mbox{.}(2025b)]%
        {Zhang2025NoteLLM2}
\bibfield{author}{\bibinfo{person}{Chao Zhang}, \bibinfo{person}{Haoxin Zhang}, \bibinfo{person}{Shiwei Wu}, \bibinfo{person}{Di Wu}, \bibinfo{person}{Tong Xu}, \bibinfo{person}{Xiangyu Zhao}, \bibinfo{person}{Yan Gao}, \bibinfo{person}{Yao Hu}, {and} \bibinfo{person}{Enhong Chen}.} \bibinfo{year}{2025}\natexlab{b}.
\newblock \showarticletitle{NoteLLM-2: Multimodal Large Representation Models for Recommendation}. In \bibinfo{booktitle}{\emph{Proceedings of the 31st ACM SIGKDD Conference on Knowledge Discovery and Data Mining V.1}} (Toronto ON, Canada) \emph{(\bibinfo{series}{KDD '25})}. \bibinfo{publisher}{Association for Computing Machinery}, \bibinfo{address}{New York, NY, USA}, \bibinfo{pages}{2815–2826}.
\newblock
\showISBNx{9798400712456}
\href{https://doi.org/10.1145/3690624.3709440}{doi:\nolinkurl{10.1145/3690624.3709440}}


\bibitem[Zhang et~al\mbox{.}(2024a)]%
        {zhang2024prospectpersonalizedrecommendationlarge}
\bibfield{author}{\bibinfo{person}{Jizhi Zhang}, \bibinfo{person}{Keqin Bao}, \bibinfo{person}{Wenjie Wang}, \bibinfo{person}{Yang Zhang}, \bibinfo{person}{Wentao Shi}, \bibinfo{person}{Wanhong Xu}, \bibinfo{person}{Fuli Feng}, {and} \bibinfo{person}{Tat-Seng Chua}.} \bibinfo{year}{2024}\natexlab{a}.
\newblock \bibinfo{title}{Prospect Personalized Recommendation on Large Language Model-based Agent Platform}.
\newblock
\showeprint[arxiv]{2402.18240}~[cs.IR]
\urldef\tempurl%
\url{https://arxiv.org/abs/2402.18240}
\showURL{%
\tempurl}


\bibitem[Zhang et~al\mbox{.}(2019)]%
        {zhang2019FDSA}
\bibfield{author}{\bibinfo{person}{Tingting Zhang}, \bibinfo{person}{Pengpeng Zhao}, \bibinfo{person}{Yanchi Liu}, \bibinfo{person}{Victor~S. Sheng}, \bibinfo{person}{Jiajie Xu}, \bibinfo{person}{Deqing Wang}, \bibinfo{person}{Guanfeng Liu}, {and} \bibinfo{person}{Xiaofang Zhou}.} \bibinfo{year}{2019}\natexlab{}.
\newblock \showarticletitle{Feature-level deeper self-attention network for sequential recommendation}. In \bibinfo{booktitle}{\emph{Proceedings of the 28th International Joint Conference on Artificial Intelligence}} (Macao, China) \emph{(\bibinfo{series}{IJCAI'19})}. \bibinfo{publisher}{AAAI Press}, \bibinfo{pages}{4320–4326}.
\newblock
\showISBNx{9780999241141}


\bibitem[Zhang et~al\mbox{.}(2024c)]%
        {Zhang2024RecGPTGP}
\bibfield{author}{\bibinfo{person}{Yabin Zhang}, \bibinfo{person}{Wenhui Yu}, \bibinfo{person}{Erhan Zhang}, \bibinfo{person}{Xu Chen}, \bibinfo{person}{Lantao Hu}, \bibinfo{person}{Peng Jiang}, {and} \bibinfo{person}{Kun Gai}.} \bibinfo{year}{2024}\natexlab{c}.
\newblock \showarticletitle{RecGPT: Generative Personalized Prompts for Sequential Recommendation via ChatGPT Training Paradigm}.
\newblock \bibinfo{journal}{\emph{ArXiv}}  \bibinfo{volume}{abs/2404.08675} (\bibinfo{year}{2024}).
\newblock
\urldef\tempurl%
\url{https://api.semanticscholar.org/CorpusID:269148580}
\showURL{%
\tempurl}


\bibitem[Zhang et~al\mbox{.}(2025a)]%
        {Zhang2025LLMPoweredUserSim}
\bibfield{author}{\bibinfo{person}{Zijian Zhang}, \bibinfo{person}{Shuchang Liu}, \bibinfo{person}{Ziru Liu}, \bibinfo{person}{Rui Zhong}, \bibinfo{person}{Qingpeng Cai}, \bibinfo{person}{Xiangyu Zhao}, \bibinfo{person}{Chunxu Zhang}, \bibinfo{person}{Qidong Liu}, {and} \bibinfo{person}{Peng Jiang}.} \bibinfo{year}{2025}\natexlab{a}.
\newblock \showarticletitle{LLM-powered user simulator for recommender system}. In \bibinfo{booktitle}{\emph{Proceedings of the Thirty-Ninth AAAI Conference on Artificial Intelligence and Thirty-Seventh Conference on Innovative Applications of Artificial Intelligence and Fifteenth Symposium on Educational Advances in Artificial Intelligence}} \emph{(\bibinfo{series}{AAAI'25/IAAI'25/EAAI'25})}. \bibinfo{publisher}{AAAI Press}, Article \bibinfo{articleno}{1483}, \bibinfo{numpages}{9}~pages.
\newblock
\showISBNx{978-1-57735-897-8}
\href{https://doi.org/10.1609/aaai.v39i12.33456}{doi:\nolinkurl{10.1609/aaai.v39i12.33456}}


\bibitem[Zhang et~al\mbox{.}(2026)]%
        {zhang2026unleashingnativerecommendationpotential}
\bibfield{author}{\bibinfo{person}{Zhiyang Zhang}, \bibinfo{person}{Junda She}, \bibinfo{person}{Kuo Cai}, \bibinfo{person}{Bo Chen}, \bibinfo{person}{Shiyao Wang}, \bibinfo{person}{Xinchen Luo}, \bibinfo{person}{Qiang Luo}, \bibinfo{person}{Ruiming Tang}, \bibinfo{person}{Han Li}, \bibinfo{person}{Kun Gai}, {and} \bibinfo{person}{Guorui Zhou}.} \bibinfo{year}{2026}\natexlab{}.
\newblock \bibinfo{title}{Unleashing the Native Recommendation Potential: LLM-Based Generative Recommendation via Structured Term Identifiers}.
\newblock
\showeprint[arxiv]{2601.06798}~[cs.IR]
\urldef\tempurl%
\url{https://arxiv.org/abs/2601.06798}
\showURL{%
\tempurl}


\bibitem[Zhao et~al\mbox{.}(2025)]%
        {zhao2025surveylargelanguagemodels}
\bibfield{author}{\bibinfo{person}{Wayne~Xin Zhao}, \bibinfo{person}{Kun Zhou}, \bibinfo{person}{Junyi Li}, \bibinfo{person}{Tianyi Tang}, \bibinfo{person}{Xiaolei Wang}, \bibinfo{person}{Yupeng Hou}, \bibinfo{person}{Yingqian Min}, \bibinfo{person}{Beichen Zhang}, \bibinfo{person}{Junjie Zhang}, \bibinfo{person}{Zican Dong}, \bibinfo{person}{Yifan Du}, \bibinfo{person}{Chen Yang}, \bibinfo{person}{Yushuo Chen}, \bibinfo{person}{Zhipeng Chen}, \bibinfo{person}{Jinhao Jiang}, \bibinfo{person}{Ruiyang Ren}, \bibinfo{person}{Yifan Li}, \bibinfo{person}{Xinyu Tang}, \bibinfo{person}{Zikang Liu}, \bibinfo{person}{Peiyu Liu}, \bibinfo{person}{Jian-Yun Nie}, {and} \bibinfo{person}{Ji-Rong Wen}.} \bibinfo{year}{2025}\natexlab{}.
\newblock \bibinfo{title}{A Survey of Large Language Models}.
\newblock
\showeprint[arxiv]{2303.18223}~[cs.CL]
\urldef\tempurl%
\url{https://arxiv.org/abs/2303.18223}
\showURL{%
\tempurl}


\bibitem[Zhao et~al\mbox{.}(2024)]%
        {Zhao2024LetMe}
\bibfield{author}{\bibinfo{person}{Yuyue Zhao}, \bibinfo{person}{Jiancan Wu}, \bibinfo{person}{Xiang Wang}, \bibinfo{person}{Wei Tang}, \bibinfo{person}{Dingxian Wang}, {and} \bibinfo{person}{Maarten de Rijke}.} \bibinfo{year}{2024}\natexlab{}.
\newblock \showarticletitle{Let Me Do It For You: Towards LLM Empowered Recommendation via Tool Learning}. In \bibinfo{booktitle}{\emph{Proceedings of the 47th International ACM SIGIR Conference on Research and Development in Information Retrieval}} (Washington DC, USA) \emph{(\bibinfo{series}{SIGIR '24})}. \bibinfo{publisher}{Association for Computing Machinery}, \bibinfo{address}{New York, NY, USA}, \bibinfo{pages}{1796–1806}.
\newblock
\showISBNx{9798400704314}
\href{https://doi.org/10.1145/3626772.3657828}{doi:\nolinkurl{10.1145/3626772.3657828}}


\bibitem[Zhong et~al\mbox{.}(2024)]%
        {zhong2024explaining}
\bibfield{author}{\bibinfo{person}{Ruiqi Zhong}, \bibinfo{person}{Heng Wang}, \bibinfo{person}{Dan Klein}, {and} \bibinfo{person}{Jacob Steinhardt}.} \bibinfo{year}{2024}\natexlab{}.
\newblock \showarticletitle{Explaining Datasets in Words: Statistical Models with Natural Language Parameters}. In \bibinfo{booktitle}{\emph{The Thirty-eighth Annual Conference on Neural Information Processing Systems}}.
\newblock
\urldef\tempurl%
\url{https://openreview.net/forum?id=u5BkOgWWZW}
\showURL{%
\tempurl}


\bibitem[Zhou et~al\mbox{.}(2020)]%
        {zhou2020s3rec}
\bibfield{author}{\bibinfo{person}{Kun Zhou}, \bibinfo{person}{Hui Wang}, \bibinfo{person}{Wayne~Xin Zhao}, \bibinfo{person}{Yutao Zhu}, \bibinfo{person}{Sirui Wang}, \bibinfo{person}{Fuzheng Zhang}, \bibinfo{person}{Zhongyuan Wang}, {and} \bibinfo{person}{Ji-Rong Wen}.} \bibinfo{year}{2020}\natexlab{}.
\newblock \showarticletitle{S3-Rec: Self-Supervised Learning for Sequential Recommendation with Mutual Information Maximization}. In \bibinfo{booktitle}{\emph{Proceedings of the 29th ACM International Conference on Information \& Knowledge Management}} (Virtual Event, Ireland) \emph{(\bibinfo{series}{CIKM '20})}. \bibinfo{publisher}{Association for Computing Machinery}, \bibinfo{address}{New York, NY, USA}, \bibinfo{pages}{1893–1902}.
\newblock
\showISBNx{9781450368599}
\href{https://doi.org/10.1145/3340531.3411954}{doi:\nolinkurl{10.1145/3340531.3411954}}


\bibitem[Zhu et~al\mbox{.}(2021)]%
        {zhu2021bars}
\bibfield{author}{\bibinfo{person}{Jieming Zhu}, \bibinfo{person}{Jinyang Liu}, \bibinfo{person}{Shuai Yang}, \bibinfo{person}{Qi Zhang}, {and} \bibinfo{person}{Xiuqiang He}.} \bibinfo{year}{2021}\natexlab{}.
\newblock \showarticletitle{Open Benchmarking for Click-Through Rate Prediction}. In \bibinfo{booktitle}{\emph{{CIKM} '21: The 30th {ACM} International Conference on Information and Knowledge Management, Virtual Event, Queensland, Australia, November 1 - 5, 2021}}, \bibfield{editor}{\bibinfo{person}{Gianluca Demartini}, \bibinfo{person}{Guido Zuccon}, \bibinfo{person}{J.~Shane Culpepper}, \bibinfo{person}{Zi~Huang}, {and} \bibinfo{person}{Hanghang Tong}} (Eds.). \bibinfo{publisher}{{ACM}}, \bibinfo{pages}{2759--2769}.
\newblock
\href{https://doi.org/10.1145/3459637.3482486}{doi:\nolinkurl{10.1145/3459637.3482486}}


\bibitem[Zhu et~al\mbox{.}(2025a)]%
        {zhu25allmbased}
\bibfield{author}{\bibinfo{person}{Lixi Zhu}, \bibinfo{person}{Xiaowen Huang}, {and} \bibinfo{person}{Jitao Sang}.} \bibinfo{year}{2025}\natexlab{a}.
\newblock \showarticletitle{A LLM-based Controllable, Scalable, Human-Involved User Simulator Framework for Conversational Recommender Systems}. In \bibinfo{booktitle}{\emph{Proceedings of the ACM on Web Conference 2025}} (Sydney NSW, Australia) \emph{(\bibinfo{series}{WWW '25})}. \bibinfo{publisher}{Association for Computing Machinery}, \bibinfo{address}{New York, NY, USA}, \bibinfo{pages}{4653–4661}.
\newblock
\showISBNx{9798400712746}
\href{https://doi.org/10.1145/3696410.3714858}{doi:\nolinkurl{10.1145/3696410.3714858}}


\bibitem[Zhu et~al\mbox{.}(2025b)]%
        {zhu-etal-2025-llm-based}
\bibfield{author}{\bibinfo{person}{Yaochen Zhu}, \bibinfo{person}{Harald Steck}, \bibinfo{person}{Dawen Liang}, \bibinfo{person}{Yinhan He}, \bibinfo{person}{Nathan Kallus}, {and} \bibinfo{person}{Jundong Li}.} \bibinfo{year}{2025}\natexlab{b}.
\newblock \showarticletitle{{LLM}-based Conversational Recommendation Agents with Collaborative Verbalized Experience}. In \bibinfo{booktitle}{\emph{Findings of the Association for Computational Linguistics: EMNLP 2025}}, \bibfield{editor}{\bibinfo{person}{Christos Christodoulopoulos}, \bibinfo{person}{Tanmoy Chakraborty}, \bibinfo{person}{Carolyn Rose}, {and} \bibinfo{person}{Violet Peng}} (Eds.). \bibinfo{publisher}{Association for Computational Linguistics}, \bibinfo{address}{Suzhou, China}, \bibinfo{pages}{2207--2220}.
\newblock
\showISBNx{979-8-89176-335-7}
\href{https://doi.org/10.18653/v1/2025.findings-emnlp.119}{doi:\nolinkurl{10.18653/v1/2025.findings-emnlp.119}}


\end{thebibliography}

\appendix
\section{Appendix}

\subsection{Implementation Details}

\begin{table*}
    \centering
    \caption{Statistics of public datasets used. “Avg.” denotes the average sequence length.}
    \label{tab:dataset_stats}
    \begin{tabular}{lcccc}
        \toprule
        \textbf{Datasets} & \textbf{\#Users} & \textbf{\#Items} & \textbf{\#Interactions} & \textbf{Avg.} \\
        \midrule
        Sports & 35,598 & 18,357 & 260,739 & 8.32 \\
        Beauty & 22,363 & 12,101 & 176,139 & 8.87 \\
        CDs & 75,258 & 64,443 & 1,022,334 & 14.58 \\
        \bottomrule
    \end{tabular}
\end{table*}

\begin{table}[h]
\centering
\caption{Hyperparameter settings for each dataset.}
\label{tab:hyperparameters}
\begin{tabular}{@{}lccc@{}}
\toprule
\textbf{Hyperparameter} & \textbf{Sports} & \textbf{Beauty} & \textbf{CDs} \\
\midrule
learning\_rate & 0.005 & 0.005 & 0.01 \\
n\_semid & 6+1 & 3+1 & 6+1 \\
vocab\_size & 2487+80 & 1103+1145 & 742+192 \\
warmup\_steps & 10,000 & 10,000 & 10,000 \\
dropout\_rate & 0.1 & 0.1 & 0.1 \\
weight\_decay & 0.07 & 0.10 & 0.07 \\
beam\_size & 50 & 50 & 50 \\
num\_layers & 4 & 4 & 4 \\
d\_model & 128 & 128 & 256 \\
d\_ff & 1,024 & 1,024 & 2,048 \\
num\_heads & 6 & 6 & 6 \\
d\_kv & 64 & 64 & 64 \\
optimizer & adamw & adamw & adamw \\
lr\_scheduler & cosine & cosine & cosine \\
train\_batch\_size & 256 & 256 & 256 \\
max\_epochs & 200 & 200 & 200 \\
early\_stop\_patience & 20 & 20 & 20 \\
\bottomrule
\end{tabular}
\end{table}

In this section, we cover our implementation details on public benchmark experiments disclosed in this work.

\subsubsection{Baselines and pre-processing.} We directly reuse the data processing implementation open-sourced by~\citealt{hou2025actionpiece} for fair and straightforward comparison.
We reuse the reported baseline performance in~\citealt{hou2025actionpiece}, where baseline results for BERTRec, SASRec, FDSA, and S³-Rec on the Sports and Beauty domains are taken directly from each respective work, and the rest are re-implementations by~\citeauthor{hou2025actionpiece}.

\subsubsection{Hyperparameter for GR Training.} We train the generative recommender model from scratch with \ourmethodauto as semantic IDs for a maximum of 200 epochs, breaking the training if there are no improvements in $N@10$ for 20 epochs.
Following~\citealt{hou2025actionpiece}, we set training batch size to 256 and tune learning rate from $\{1 \times 10^{-3}, 3 \times 10^{-3}, 5 \times 10^{-3}\}$ and weight-decay from $\{0.07, 0.1, 0.15, 0.2\}$.
We conduct three repeated experiments with random seeds $\{2025, 2026, 2027\}$, and report the test set performance with the model checkpoints with best average $N@10$ across runs.

\subsubsection{GR Inference.} We use standard auto-regressive decoding with a beam size of 20 for inference. 
However, since \ourmethodauto can produce variable length semantic IDs, we do not sample from the base generative recommender for a fixed number of steps; instead, we generate until we reach the end-of-sequence token.
The generated sequences are then ranked by their beam scores.

\subsubsection{\ourmethodauto} By default, we set the initial cluster size to 15 across all datasets and the number of refinement cycles to 3 (while the architect LLM can choose to exceed this number). 
The minimum cluster size for sub-clustering is set to 30, i.e., any group of more than 30 items would trigger the procedure for producing sub-category descriptors for the said group.
We break the iterative refinement loop if the item coverage for any cluster is greater than 95 or if the refiner LLM fails to generate more than 20 error cases. 
Due to the large number of items in the CDs domain, we set the sub-clustering threshold to 100 for the branching-controlled version of the algorithm and 500 for the subsampled variant.
While the final vocabulary size of \ourmethodauto varies due to the stochasticity of program execution, we report the base vocabulary and the number of conflict resolvers (i.e., resolving tokens used when two items hash to the same set of \ourmethodauto tags) as ``base tokens + resolvers'' in~\Cref{tab:hyperparameters} under ``vocab\_size.'' 
As shown, \ourmethodauto is significantly more efficient than its semantic ID counterparts, using fewer than 3,000 words across the board on academic benchmarks.

\section{Further Related Works}

\noindent\textbf{ML Models with Natural Language Parameters.} 
Beyond recommender systems, there have been continuing efforts to introduce parameters tied to explicit natural language concepts into ML models, e.g., concept-bottleneck models~\cite{Koh2020ConceptBM, sun2025concept}, typically for their improved interpretability~\cite{zhong2024explaining}.
However, this stream of work generally does not aim for improved accuracy and frequently assumes there exist sets of concepts (in our case, features) that are readily available~\cite{Koh2020ConceptBM}.
The most relevant work to ours is perhaps the label-free concept bottleneck model~\cite{sun2025concept}, which prompts an LLM with a task description for concepts.
However, our aim is to infer a set of features from large sets of items for recommendation, the nature of which is typically difficult to infer from the task description itself, unlike domains such as image classification with a fixed set of labels.\\

\noindent\textbf{LLM for Text Mining At Scale} 
Another relevant stream of work to \ourmethod, as discussed throughout the paper, consists of recent advances in leveraging LLMs for scalable text mining, typically manifesting as topic/latent theme mining~\cite{xie-etal-2025-latent, pham-etal-2024-topicgpt} or taxonomy mining~\cite{Wan2024TnTLLM}. 
However, these frameworks generally do not consider downstream applications and usually impose sequential dependencies on the interaction between LLMs and text corpora~\cite{Wan2024TnTLLM, pham-etal-2024-topicgpt}, whereas \ourmethod uses parallelized operations, making scaling to recommendation-scale item corpora feasible.\\

\noindent\textbf{LLM-generated Semantic IDs}
Concurrent to this work, we note there are also recent explorations of LLM-generated natural language semantic IDs, dubbed Term IDs~\cite{zhang2026unleashingnativerecommendationpotential}, where one prompts an LLM for free-form generated semantic IDs over a batch of similar items for adapting pre-trained LLMs for recommendation. 
To this end, Term IDs impose fewer constraints on the vocabulary itself, since a pre-trained LLM is able to generalize information across semantically similar Term IDs.
On the other hand, \ourmethodauto aims to develop a descriptor vocabulary with minimal constraints on the downstream model.
To illustrate this point, \ourmethodauto enables a 2-layer transformer model (the same as Tiger~\cite{rajput2023recommender}) to perform better than the 0.4B variant of LLM-based GR with Term IDs on the Sports and Beauty domains, showcasing the effectiveness of \ourmethodauto.

\subsection{Sample Prompts (Sports Domain)}
\label{subsec:sample_prompts_sports}

\begin{promptbox}[Architect - Generate Inital Vocabulary]
\ttfamily 
You are an expert taxonomy architect. Your task is to create a set of fine-grained sub-categories.\\
The items you are categorizing all belong to the parent category: "\textcolor{blue}{\{parent\_rule\_description\}}".\\
\\
Your new categories MUST be more specific subdivisions of this parent category. Do not simply repeat the parent category's name. For example, if the parent is "Firearm Accessories", your sub-categories should be "Optics", "Holsters", "Cleaning Kits", etc., not "Firearm Accessories" again.\\
\\
Primary Goal: Create a set of \~{}\textcolor{blue}{\{n\_target\_rules\}} mutually exclusive sub-categories that logically partition the parent.\\
\\
\textcolor{blue}{\{context\_prompt\}}\\
Input Data:\\
Below are diverse product examples. Use them to understand the product landscape you need to partition.\\
\textcolor{blue}{\{sample\_text\}}\\
\\
Task \& Strict Output Format:\\
Devise a set of categories. For each, provide a "name" and a "description" with "INCLUDES" and "EXCLUDES" clauses. Respond ONLY with a single JSON object: \{\textcolor{blue}{"categories": [\{...\}]}\}.\\
\\
Your JSON Response:
\end{promptbox}

\begin{promptbox}[Annotator]
\ttfamily 
\textcolor{blue}{\{context\_prompt\}}\\
Above, I have provided the parent category and some product examples.\\
\\
Your Goal:\\
Propose a list of exactly \textcolor{blue}{\{n\_target\_rules\}} sub-categories.\\
\\
Constraint - Overlap:\\
Sub-categories must be mutually exclusive.\\
\\
Constraint - Coverage:\\
Together, they should cover most common items found in "\textcolor{blue}{\{parent\_rule\_description\}}".\\
\\
Constraint - Specificity:\\
Do NOT create a "Miscellaneous" or "Other" category. Every category must have a specific, meaningful name.\\
\\
Constraint - Style:\\
Use professional, industry-standard terminology.\\
\\
Output Format:\\
Your response must be a JSON object with the following structure:\\
\{\\
\ \ "categories": [\\
\ \ \ \ \{\\
\ \ \ \ \ \ "name": "Category Name",\\
\ \ \ \ \ \ "description": "Definition of what is included",\\
\ \ \ \ \ \ "includes": ["Example 1", "Example 2"],\\
\ \ \ \ \ \ "excludes": ["Non-example 1"]\\
\ \ \ \ \}\\
\ \ ]\\
\}\\
\end{promptbox}

\begin{promptbox}[Annotator - Generating Error Feedbacks]
\ttfamily 
You are a data analyst specializing in taxonomy quality control. Analyze a cluster of \textcolor{blue}{\{len(ticket\_cluster)\}} uncategorized products and synthesize a single, structured change proposal.\\
\\
Existing Category Rules:\\
\textcolor{blue}{\{existing\_rules\_text\}}\\
\\
Analysis of Uncategorized Items:\\
\textcolor{blue}{\{ticket\_examples\_text\}}\\
\\
Task: Decide the most logical action and formulate a proposal. Respond ONLY with a single JSON object. Your response MUST strictly follow one of the formats below.\\
\\
1. To create a new category:\\
\{"change\_type": "CREATE\_NEW\_CATEGORY", "problem\_summary": "<summary>", "suggested\_change": \{"new\_rule\_description": "New Category Name: INCLUDES: ... EXCLUDES: ..."\}\}\\
\\
2. To expand an existing category (THIS IS A VERY STRICT FORMAT):\\
The "refined\_description" MUST be a single complete string, starting with a name, then "INCLUDES", then "EXCLUDES".\\
GOOD EXAMPLE:\\
\{\\
\ \ "change\_type": "EXPAND\_EXISTING\_CATEGORY",\\
\ \ "problem\_summary": "The current 'Outdoor Gear' rule is too generic and should explicitly include tactical accessories.",\\
\ \ "suggested\_change": \{\\
\ \ \ \ "rule\_id\_to\_refine": "rule\_a4368cef",\\
\ \ \ \ "refined\_description": "Outdoor \& Tactical Gear: INCLUDES: Tents, backpacks, sleeping bags, and tactical accessories like gloves, belts, and pouches. EXCLUDES: Specialized sporting equipment, firearms, and knives."\\
\ \ \}\\
\}\\
\\
3. To ignore outliers:\\
\{"change\_type": "IGNORE\_AS\_OUTLIERS", "problem\_summary": "<summary>", "suggested\_change": \{"reason": "<reason>"\}\}\\
\\
Your JSON Response:
\end{promptbox}

\begin{promptbox}[Architect: Taxonomy Refinement]
\ttfamily 
You are a senior taxonomy manager. Review the following change proposals and decide whether to approve or reject each. Your goal is a clean, non-overlapping taxonomy. But if the proposal is reasonable, you should try to accomodate that. You should also reject proposals that are out-of-place by common sense. Respond ONLY with a JSON list of objects, each with "proposal\_id", "decision" ('APPROVED' or 'REJECTED'), and "reasoning".\\
\\
Proposals for Review:\\
\textcolor{blue}{\{proposals\_text\}}\\
\\
Your JSON Decision List:
\end{promptbox}

\subsection{Ranking Performance on Smaller Dataset}
\label{subsec:ranking_performance_smaller_dataset}

\begin{table}[h]
\centering
\caption{Ranking performance comparison between Item ID baseline and \ourmethod.}
\label{tab:ranking_results_small}
\begin{tabular}{lcc}
\toprule
\textbf{Metric} & \textbf{Item ID (Baseline)} & \textbf{\ourmethod (Ours)} \\
\midrule
R@5  & 0.0833 $\pm$ 0.0068 & \textbf{0.1205 $\pm$ 0.0080} \\
N@5    & 0.0589 $\pm$ 0.0051 & \textbf{0.0853 $\pm$ 0.0060} \\
\midrule
R@10 & 0.1163 $\pm$ 0.0079 & \textbf{0.1751 $\pm$ 0.0093} \\
N@10   & 0.0695 $\pm$ 0.0052 & \textbf{0.1028 $\pm$ 0.0061} \\
\bottomrule
\end{tabular}
\end{table}

In this section, we discuss further performance comparisons in ranking experiments, where we report both Recall and NDCG in~\Cref{tab:ranking_results_small}. 
This subset contains 1,791 users, 15,830 items, and 47,260 interactions, making the collaborative filtering signal less salient due to the smaller dataset size.
As shown, our method consistently outperforms the ID-based method by a larger margin than in the main experiment, verifying our intuition: when the dataset is large and the collaborative filtering signal is more stable, assigning more item-specific parameters helps in memorizing item-specific information, sometimes leading to better ranking performance. 
Conversely, when the dataset is sparse, content-aware structured discrete features shared among items help convergence significantly, making \ourmethod features potentially applicable to cold-start settings, where they lead to better generalization.

\subsection{Cardinality of Free-form Generation}
\label{subsec:vocab_ultilization}

To investigate whether free-form generation indeed leads to vocabulary explosion, we conduct an ablation study where we prompt an LLM to generate three keywords over the private dataset discussed in~\Cref{sec:experiments}. 
Indeed, there were approximately 90k unique keywords generated over 210k documents; however, only 21k (or 24\%) of the descriptors appeared in more than one item. 
Consequently, these features are inapplicable for natively acting as compact, ID-like features, making downstream modeling challenging. 
On the other hand, descriptors generated by \ourmethod have a much higher utilization rate (87\%) across a vocabulary of over 2k, demonstrating their effectiveness.

\subsection{Tags Generated}
\label{subsec:generated_tags}

We show qualitative examples of the discovered vocabulary in~\Cref{tab:cross-domain_vocabulary}. 
For presentation purposes, we pick two descriptors per level but only expand on one of them until no valid children are found.
Note that in \ourmethod, the depth of each hierarchy path can be of variable length, as shown in the CDs domain example.
As showcased in the table, \ourmethod is able to find clear separations of products, especially towards the earlier layers of the hierarchy (e.g., ``Skincare'' and ``Haircare'' products). 
However, towards the end of the hierarchy, the boundary between products becomes fuzzy, and it is more challenging to come up with clean, disentangled separation rules (e.g., ``Specialty Function'' and ``Ceramic Plate Irons''), making it harder to enforce a nominal constraint on features. 
While we partially address this issue by instructing the LLM to select the best-fitting descriptor for each product when multiple are technically correct, developing methods to fundamentally address such disentanglement issues remains an open problem. 
Nevertheless, even at later layers, \ourmethod can still discover meaningful descriptors that separate the products, such as sleeping bags by their applicable temperature range, as shown in the Sports domain.

\subsection{Effect of Multi-agent Self-refinement on Other Domains}
\label{subsec:coverage_impv_across_domains}

\begin{figure*}[t]
    \centering
    \includegraphics[width=0.32\textwidth]{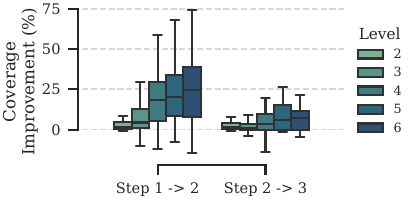} \hfill
    \includegraphics[width=0.32\textwidth]{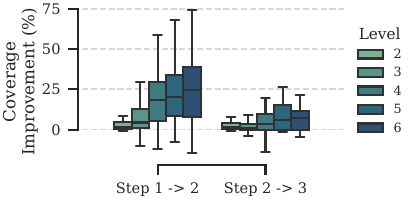} \hfill
    \includegraphics[width=0.32\textwidth]{assets/sports_refinement_improvement.pdf}
    
    \caption{Coverage improvements across domains. We discuss further details in~\Cref{subsec:coverage_impv_across_domains}.}
    \label{fig:horizontal_comparison_coverage_impv}
\end{figure*}

We showcase the improvements in coverage by reporting the distribution of delta changes in the coverage rate in \Cref{fig:horizontal_comparison_coverage_impv}.
The trend is consistent across domains: (1) our designed multi-agent self-refinement loop effectively refines the vocabulary, improving item coverage; and (2) the trend of improvement (i.e., delta greater than zero) becomes less stable in later layers, possibly because it is more challenging to generate disentangled descriptors, as qualitatively analyzed in \Cref{subsec:generated_tags}.
Regardless, the Wilcoxon signed-rank test shows that improvements are significant (i.e., delta greater than zero) with p-values smaller than 0.05, except for levels five and six in the CDs domain, corroborating the general effectiveness of the self-refinement mechanism.

\begin{table*}[t]
\small
\centering
\caption{Discovered Vocabulary (Sub-sampled) Across Domains}
\label{tab:cross-domain_vocabulary}
\begin{tabular}{p{5.2cm} p{5.2cm} p{5.2cm}}
\toprule
\textbf{Beauty Domain} & \textbf{Music (CDs) Domain} & \textbf{Sports Domain} \\
\midrule
L1: Skincare - Facial Treatments & L1: Jazz Music & L1: Fitness \& Exercise Equipment \\
L1: Haircare - Treatments \& Styling & L1: Electronic Dance Music (EDM) & L1: Camping \& Hiking Gear \\
\quad L2: Hair \& Scalp Treatments & \quad L2: Smooth Jazz & \quad L2: Shelter Systems \\
\quad L2: Hair Styling Tools \& Acc. & \quad L2: Free Jazz \& Avant-Garde & \quad L2: Sleeping Systems \\
\quad\quad L3: Hair Dryers & \quad\quad L3: Interdisciplinary Avant-Garde & \quad\quad L3: Sleeping Bag Liners \\
\quad\quad L3: Flat Irons \& Straighteners & \quad\quad L3: Free Improv. \& Avant-Garde & \quad\quad L3: Sleeping Bags \\
\quad\quad\quad L4: Travel \& Mini Flat Irons & \quad\quad\quad L4: Free Jazz (Post-Bebop) & \quad\quad\quad L4: Sleeping Bag Shapes \\
\quad\quad\quad L4: Professional Salon Irons & \quad\quad\quad L4: Avant-Garde (Exp.) & \quad\quad\quad L4: Temp.-Rated Bags \\
\quad\quad\quad\quad L5: Specialty Function & \quad\quad\quad\quad L5: Atonal Exploration & \quad\quad\quad\quad L5: Mild to Cool Bags \\
\quad\quad\quad\quad L5: Ceramic Plate Irons & \quad\quad\quad\quad L5: Genre-Infused Exp. & \quad\quad\quad\quad L5: Cold Weather Bags \\
\quad\quad\quad\quad\quad L6: Adv. Plate Tech. & & \quad\quad\quad\quad\quad L6: 10-30 Degree Bags \\
\quad\quad\quad\quad\quad L6: Ionic Technology & & \quad\quad\quad\quad\quad L6: 0-Degree Rated Bags \\
\bottomrule
\end{tabular}
\end{table*}

\begin{table*}[t]
\caption{Summary of Notations \label{tab:notations}}
\centering 
\small 
\begin{tabular}{l p{0.7\columnwidth}} 
    \toprule
    Notation & Description \\ 
    \midrule
    $\mathcal{U}, \mathcal{I}$ & Sets of users and items \\
    $S_u$ & Interaction history for user $u \in \mathcal{U}$ \\
    $X_i$ & Raw attributes for item $i \in \mathcal{I}$ \\
    $\Pi, \pi$ & LLMs, and a specific instance \\
    $\mathcal{E}$ & Set of natural language error feedback \\
    $A$ & Descriptor feature vocabulary \\
    $a$ & A specific feature descriptor ($a \in A$) \\
    $g(\cdot)$ & Transformation mapping attributes to a subset of $A$ \\
    $f_{\theta}$ & Recommender system parameterized by $\theta$ \\
    \bottomrule
\end{tabular}
\end{table*}

\begin{table*}[t]
\centering
\caption{Impact of the backbone model used in \ourmethod on downstream performance, Recall and NDCG.}
\label{tab:ablation_backbones}
\setlength{\tabcolsep}{4pt} 
\small
\begin{tabular}{@{}l cccc cccc cccc@{}}
\toprule
\multirow{2}{*}{\textbf{Backbone LLM}} & \multicolumn{4}{c}{\textbf{Sports}} & \multicolumn{4}{c}{\textbf{Beauty}} & \multicolumn{4}{c}{\textbf{CDs}} \\
\cmidrule(lr){2-5} \cmidrule(lr){6-9} \cmidrule(lr){10-13}
 & R@5 & N@5 & R@10 & N@10 & R@5 & N@5 & R@10 & N@10 & R@5 & N@5 & R@10 & N@10 \\
\midrule
Gemini-1.5 & 0.0284 & 0.0184 & 0.0448 & 0.0237 & 0.0469 & 0.0319 & 0.0710 & 0.0397 & 0.0531 & 0.0349 & 0.0812 & 0.0439 \\
Gemini-2.5 & \textbf{0.0299} & \textbf{0.0194} & \textbf{0.0474} & \textbf{0.0251} & \textbf{0.0492} & \textbf{0.0332} & \textbf{0.0728} & \textbf{0.0408} & \textbf{0.0544} & \textbf{0.0358} & \textbf{0.0836} & \textbf{0.0453} \\
\bottomrule
\end{tabular}
\end{table*}

\end{document}